\documentclass[
prd,%twocolumn,
,showpacs,amssymb,superscriptaddress,aps,nofootinbib
]{revtex4-2}
\usepackage{graphicx}
\usepackage{color}
\input{epsf}

\usepackage{amsmath,amssymb}
\usepackage{bm}
\usepackage{times}
\usepackage{ulem}
\usepackage[colorlinks=true]{hyperref}
\hypersetup{colorlinks=true,
            citecolor=blue,
            linkcolor=red,
            urlcolor=red}
\definecolor{maroon}{cmyk}{0,0.87,0.68,0.32}

\newcommand{\dalm}{\kern1pt\vbox{\hrule height 0.9pt\hbox{\vrule width 0.9pt
\hskip 2.5pt\vbox{\vskip 5.5pt}\hskip 3pt\vrule width 0.3pt}\hrule height 0.3pt}
\kern1pt}

\begin{document}

%\twocolumn[\hsize\textwidth\columnwidth\hsize\csname @twocolumnfals\endcsname

% For two column
%\wideabs{

\title{Couplings of torsional and shear oscillations in a neutron star crust}

\author{Hajime Sotani}
\email{sotani@yukawa.kyoto-u.ac.jp}
\affiliation{Astrophysical Big Bang Laboratory, RIKEN, Saitama 351-0198, Japan}
\affiliation{Interdisciplinary Theoretical \& Mathematical Science Program (iTHEMS), RIKEN, Saitama 351-0198, Japan}
\affiliation{Theoretical Astrophysics, IAAT, University of T\"{u}bingen, 72076 T\"{u}bingen, Germany}

\author{Arthur G. Suvorov}
\affiliation{Theoretical Astrophysics, IAAT, University of T\"{u}bingen, 72076 T\"{u}bingen, Germany}
\affiliation{Departament de F{\'i}sica Aplicada, Universitat d'Alacant, Ap. Correus 99, E-03080 Alacant, Spain}
\affiliation{Manly Astrophysics, 15/41-42 East Esplanade, Manly, NSW 2095, Australia}

\author{Kostas D. Kokkotas}
\affiliation{Theoretical Astrophysics, IAAT, University of T\"{u}bingen, 72076 T\"{u}bingen, Germany}

\date{\today}

% Abstract
\begin{abstract}
Mature neutron stars are thought to be sufficiently cold that nuclei in the outer layers freeze, solidifying a crust. Crustal elasticity allows the star to support a set of seismic modes, such as torsional oscillations. These axial-parity modes can couple to the polar sector in a number of ways, for example via rotation or a magnetic field. Even in a static, spherically-symmetric star however these modes can couple at non-linear order. In this study, such couplings in the crust are examined for the first time: we derive the axisymmetric perturbation equations for second-order axial eigenfunctions, which are sourced by axial-polar couplings at first-order, and solve the resulting equations in the time domain. Through our studies, we find that the second-order spectrum contains additional oscillation modes, not predicted by the linear analysis with either axial or polar perturbations, which can be excited to relatively large amplitudes.

\end{abstract}

%
%%%%%%%%%%%%%%%%%%%%%%%%%%%%%%%%%%%%%%%%%%%%%%%%%
%  04.30.-w  :  Gravitational waves: theory
%  04.40.Dg :  Relativistic stars: structure, stability, and oscillations (see also 97.60.-s Late stages of stellar evolution) 
%  21.60.-n  :   Nuclear structure models and methods
%  21.65.Ef  :  Symmetry energy
%  26.60.+c  :  Nuclear matter aspects of neutron stars
%  26.60.Gj  :  Neutron star crust
%  97.10.Sj  :   Pulsations, oscillations, and stellar seismology 
%  97.60.Jd  :   Neutron stars (see also 26.60.+c Nuclear matter aspects of neutron stars in nuclear physics)
%%%%%%%%%%%%%%%%%%%%%%%%%%%%%%%%%%%%%%%%%%%%%%%%%
%]
% For two column
%}
\maketitle

%\baselineskip 24pt

%%%%%%%%%%%%%%%%%%%%%%%%%%%%%%%%%%%%%%%%%%%%%%%%
\section{Introduction}
\label{sec:I}
%%%%%%%%%%%%%%%%%%%%%%%%%%%%%%%%%%%%%%%%%%%%%%%%
Neutron stars, produced in supernova explosions ending the lives of massive stars, are some of the most suitable objects to probe physics at extreme scales. Their cores readily reach densities exceeding that of nuclear saturation while maintaining baryonic degrees of freedom at low temperatures, placing them in a parameter space inaccessible to terrestrial laboratories \cite{bas18}. The elastic, solid crust of a neutron star is also thought to host various exotic states of matter, such as hadron-quark mixed ``pasta'' phases \cite{rav83,hash84} and ion lattices soaked in superfluid neutrons \cite{cham06,and21}. With the exception of gravitational waves, the crust effectively mediates all of the persistent and transient activity from these stars, and thus detailed modelling of both dynamical processes and the crust itself is necessary to fully utilize existing and upcoming instruments.

The crust can sustain a variety of seismic oscillation modes \cite{Schumaker83,cham08,McDermott,Vavoul2008}. Overstraining events, particularly in the magnetar class of objects where an ultrastrong field is capable of exerting great stress \cite{rud91,land15,suv23}, might excite crust-localized \cite{dunc98,piro05} and global magnetoelastic \cite{lev07a,lev07} modes. This theoretical prediction is supported by observations of quasi-periodic oscillations (QPOs) with frequencies on the order of tens of Hz and up in the X-ray tails of magnetar flares, as in the 1979 event in SGR 0526-66 \cite{barat83}, the 1998 event in SGR 1900+14 \cite{sw05}, and the 2004 event in SGR 1806-20 \cite{is05,ws06}. The latter demonstrated an especially rich spectrum; studies coming out over a decade later are still finding new Fourier peaks \cite{ham11,hup14,mill19}. In principle, the inverse problem of deducing the stellar properties --- magnetic geometry, equation of state (EOS), crustal shear modulus, ... --- can be tackled by matching the observed spectra to detailed models. Such an approach has been carried out by various authors \cite{col11,col12,SKS07,sot12,gab16,sot19,sot24}, who matched the forest of low- and high-frequency QPOs from various flares to the modes of a general-relativistic star, with a realistic EOS and topologically-mixed, globally-threading magnetic field, exhibiting coupled axial and polar perturbations\footnote{ 
We note that the axial and polar perturbations are classified by their parity. That is, considering a coordinate transform of polar angles $\theta \to \pi-\theta$ and $\phi \to \pi+\phi$, an axial quantity of angular order $\ell$ transforms as $(-1)^{\ell+1}$ while polar ones change as $(-1)^\ell$.}.

Despite these successes, there remains a number of theoretical challenges of varying severity. Arguably the most serious of these relates to how the failure mechanism actually operates in the crust. The seed for the main event is the evolution of the magnetic field, which builds up mechanical stresses that culminate in a failure event and the release of magnetoelastic energy \cite{dunc92,thom95}. This localized relaxation generates horizontal currents that may go on to launch chains of Hall and/or thermoplastic waves that instigate further failures of various amplitude away from the initial site, twisting the external field and priming it for explosive reconnection \cite{lyu03,bel14,li16,land23}. Alternatively, the rapid release of poloidal energy that fuels the flare could trigger instability and magnetic reconfiguration \cite{suv23}, exerting Maxwell stress over short timescales. This may help to explain why the dynamic spectra of flare QPOs are so complicated: different regions of the crust fail at different times, leading to the rising and falling of various mode families over the course of hundreds of seconds. This means that a fully-fledged attempt at a solution of the aforementioned inverse problem requires not only the specification of initial data at ignition, but also injections of energy at later times, the particulars of which require detailed, microphysical calculations of failure propagation that are presently out of reach.

A second, related problem is that most studies of magnetar oscillations work within the linear regime (though see \cite{Lasky11,Zink12, Ciolfi11,Ciolfi12,gab13,leung22} for some exceptions). That is, one solves for a small perturbation over a fixed (or dynamical) background in either the time or frequency domain. While this provides a reasonable, leading picture for the spectrum of possible excitations, it does not allow one to study the physics related to saturation via parametric and other instabilities, which are important since the observed QPOs have significances that vary by orders of magnitude \cite{ws06,ham11,mill19}; the respective Bayes factors are presumably tied to the amplitude that individual modes achieve. Aside from predicting saturation amplitudes, nonlinear models are needed to deduce which daughter modes can be excited from some initial combination of parents. An example of this is the so-called $p$-$g$ instability, where nonresonant couplings between these polar modes could instigate orbital energy losses in a binary inspiral, possibly leading to significant dephasing in a gravitational waveform \cite{reed16}. In the case of QPOs in flare tails, there could be instances of frequency drifting also. For the SGR 1806-20 giant flare, Miller et al. \cite{mill19} further report a $\approx 28.1$~Hz oscillation starting at $\approx 193$~s post flare and subsequently disappearing, with a new peak at $\approx 27.9$~Hz emerging some $\gtrsim 20$~s later. Field evolution via Hall avalanches or backreaction from excited modes \cite{gab14}, nonlinear effects both, could theoretically cause such a drift (see also Ref.~\cite{lev07}). 

In this paper, which we intend to be the first in a series, we make some additional steps toward the nonlinear modelling of crustal oscillations in a neutron star. We derive equations at (Eulerian) second-order for axial perturbations in general relativity with the Cowling approximation, including new expressions for the shear and strain tensors, building on the linear formalism of Schumaker and Thorne \cite{Schumaker83}. The second-order perturbations are sourced by first-order, coupled polar-axial perturbations, and the equations are solved for a realistic EOS in the time domain. Although nonlinear torsional oscillations have been studied previously (notably in Ref.~\cite{gab13}), sourcing by axial-polar primaries has not. Our scheme is further capable of handling seed injections at later times, to simulate delayed and non-local failings within the crust, and is stable over long (many seconds) timescales. While we take a step back from the realistic picture by not considering crust-core couplings or magnetic fields, assuming that the stellar magnetic field is not so strong that one can neglect such couplings, i.e., the field strength is $\ll 10^{15}$~G \cite{gab18,SKS24}, certain key effects, which are supported by analytic estimates, are observed in the simulations. For instance, we predict the existence of new pairs of daughter modes that arise from axial-polar couplings at linear order, the amplitude of which is set by that of the parents. Higher quantum-number modes and overtones from the linear spectrum are also excited with a complicated pattern. The main goal here is to derive the relevant equations and solve them in a simplified setting to pave the way for more realistic investigations in forthcoming work.

This paper is organized as follows. 
In Sec.~\ref{sec:background}, we briefly introduce the (tabular EOS) equilibrium neutron star model adopted in this study and the shear modulus characterizing the crust elasticity. In Sec.~\ref{sec:perturbations}, we derive the two-dimensional perturbation equations for 2nd order axial oscillations through an order-matching procedure, where the 2nd order perturbation equation has a source term composed of the product of linear axial and polar perturbations. The boundary conditions imposed at the both edges of the (elastic) crust region together with the initial conditions are also described. Then, we show several examples of mode excitation by solving, in the time domain, the derived equation in Sec.~\ref{sec:2nd}. Finally, in Sec.~\ref{sec:Conclusion} we offer some conclusions. 
In this study, we adopt geometric units with $c=G=1$ for the speed of light, $c$, and the gravitational constant, $G$. The metric signature is $(-,+,+,+)$.

%%%%%%%%%%%%%%%%%%%%%%%%%%%%%%%%%%%%%%%%%%%%%%%%
\section{Equilibrium models}
\label{sec:background}
%%%%%%%%%%%%%%%%%%%%%%%%%%%%%%%%%%%%%%%%%%%%%%%%

Since we are ultimately interested in developing a scheme that applies to Galactic magnetars, which spin very slowly, the metric and fluid variables can be taken as static to a good approximation.  The background neutron star model is further taken to be spherically symmetric and strain-free, whose line element is given in Schwarzschild-like coordinates $\{t,r,\theta,\phi\}$ through
\begin{equation}
  ds^2 = -e^{2\Phi}dt^2 + e^{2\Lambda}dr^2 + r^2\left(d\theta^2 + \sin^2\theta d\phi^2\right). \label{eq:metric}
\end{equation}
With this expression, the only non-zero component of the four-velocity is in the $t$-direction, and is given by
\begin{equation}
  u^t = e^{-\Phi}.
\end{equation} 
The metric variables $\Phi$ and $\Lambda$ appearing in Eq.~\eqref{eq:metric} together with the fluid variables, $p$ (pressure) and $\varepsilon$ (energy density), are constructed by integrating the Tolman-Oppenheimer-Volkoff equations, assuming an approximate but appropriate EOS for neutron star matter, i.e., zero-temperature matter in beta-equilibrium. In this study, we specifically adopt the SLy4, which is a Skyrme-type EOS with a self-consistent crustal region \cite{SLy4}, and focus on stars with mass $M=1.41M_\odot$ and radius $R=11.68$~km. 

Since torsional and shear oscillations can be excited only inside regions with elasticity, their frequencies depend on the density at the boundaries. The outer boundary, which is the interface between the crust and the envelope (or ``ocean''), depends strongly on the physical temperature, meaning a zero-temperature EOS cannot apply~\cite{GPE83}. In this study, we simply assume that the transition (baryon) density at this interface is $10^{10}$ g/cm$^3$ and the density matching to the stellar atmosphere is $10^6$ g/cm$^3$. Meanwhile, the inner crust-core boundary is also complicated. In a realistic neutron star, it is thought that the crust region is composed not only of phases of the more-familiar spherical nuclei but also of non-spherical nuclei -- the so-called pasta phases. Nevertheless, the thickness of the pasta phase is rather narrow, compared to that of spherical nuclei, e.g., \cite{sot17}. So, in this study, we only consider the phase of spherical nuclei, by assuming that the transition (baryon) density between the crust and core is $1.045\times 10^{14}$ g/cm$^3$. With these transition densities, the radii of the core and the annulus separating the crust, $R_c$, and the envelope, $R_e$, become $R_c=10.85$ km and $R_e=11.57$ km, respectively. 

The shear modulus, $\mu$, is another important quantity associated with elastically-supported oscillations. Even though there really are multiple shear moduli which depend on the shape of nuclei \cite{PP1998}, in this study we neglect the presence of pasta and assume a single contributor. For this spherical-nuclei shear modulus we simply adopt the standard, low-temperature expression proposed in Ref.~\cite{SHOII1991}, i.e.,
\begin{equation}
  \mu = 0.1194\frac{n_i(Ze)^2}{a}, \label{eq:mu}
\end{equation}
where $e$, $n_i$, $Z$, and $a$ respectively denote the elementary charge, ion number density, charge number, and Wigner-Seitz cell radius (i.e., $4\pi a^3/3 = 1/n_i$).

\subsection{Remarks on nonlinear elasticity}

The elastic solid that is the neutron star crust is in reality a complicated, chemically- and magnetically-stratified medium with non-trivial composition and temperature gradients \cite{cham08}. At linear order though, the crust (or any other solid) necessarily abides by a Hookean relationship; that is, the material is such that strain is linearly proportional to stress. The elastic stress-tensor is itself proportional to the Lagrangian displacements at linear order in a linear mode problem. At nonlinear order, however, the Hookean approximation becomes an additional assumption of the model -- some kind of ``elastic EOS'' is needed to close the system.

As described in Section 12 of Ref.~\cite{andcom21} (see also \cite{wells}), more general elasticity relationships can be explored. A set of tensors can be introduced on the (Lagrangian) matter space, from which a set of scalar invariants can be constructed which represent a type of basis of possible relationships between the shear and strain tensors. These individual terms are weighted by general-relativistic generalizations of the Lam{\'e} coefficients. Although an amount of literature is devoted to the study of the shear and bulk modulii, higher-order coefficients have not been examined in a neutron star crust. By including these additional terms though, a more general stress tensor could be built (Eq.~(12.19) in Ref.~\cite{andcom21}), which will adjust the spectrum.

To make contact with linear studies, however, we adopt a Hookean approximation throughout. Exploration of the non-linear spectra for different stress-strain relationships will be considered elsewhere.

%%%%%%%%%%%%%%%%%%%%%%%%%%%%%%%%%%%%%%%%%%%%%%%%
\section{Perturbation equations}
\label{sec:perturbations}
%%%%%%%%%%%%%%%%%%%%%%%%%%%%%%%%%%%%%%%%%%%%%%%%

In this study, we adopt the Cowling approximation, i.e., the metric perturbations are neglected so that the matter fields oscillate on a fixed spacetime. This is a reasonable approximation since even the most energetic of polar-parity modes excited in a magnetar flare are unlikely to produce strong gravitational-wave signals \cite{Lasky11,Ciolfi11,Ciolfi12}. The torsional oscillations, which are classified as axial perturbations and form our primary interest, are less energetic by orders of magnitude.

We further restrict ourselves to axisymmetric perturbations. This approximation is less physically justifiable, especially since axisymmetric states seem to be an evolutionarily-unlikely outcome in magnetars \cite{bec22,suvg23}, but it is sufficient to demonstrate the main effects of polar-axial coupling at non-linear order. In general, the spatial components\footnote{Throughout this work, purely spatial components are written with Latin scripts $(i,j,k,\ldots)$, while spacetime indices are denoted by members of the Greek alphabet $(\alpha,\beta,\gamma,\ldots)$.} of the Lagrangian displacement vector can thus be given by
%%%%
\begin{equation} \label{eq:displacement}
  \xi^i = \left(rW(t,r,\theta),V(t,r,\theta),\frac{{\cal Y}(t,r,\theta)}{\sin\theta}\right),
\end{equation}
%%%%
where $W$ and $V$ are of polar parity and ${\cal Y}$ encompasses axial perturbations. 
In what follows, perturbed variables associated with such a Lagrangian displacement are written with a prepended $\delta$, which in general encompasses a hierarchy of terms. For example, the perturbed velocity can be expanded as 
%%%%
\begin{equation}
  \delta u_\mu = \delta u_\mu^{(1)} + \delta u_\mu^{(2)} + \cdots, \\
\end{equation}
%%%%
where the superscript, $(i)$, denotes the $i$-th order perturbations of each quantity\footnote{We remark that at this point the scheme differs from several previous studies looking at saturation amplitudes \cite{dz80,pk15,pk16}, not only because of the relativistic framework but because we explicitly include a second-order velocity perturbation, which is often set to zero.}.
 
Using the displacement given by Eq.~\eqref{eq:displacement}, the spatial parts of the perturbed four-velocity can be written as
%%%%%%%%
\begin{eqnarray}
  \delta u^i &=& \left(re^{-\Phi}\partial_t W,e^{-\Phi}\partial_t V,e^{-\Phi}\frac{\partial_t{\cal Y}}{\sin\theta}\right), 
     \label{eq:du^i}  \\
  \delta u_i &=& \left(re^{-\Phi+2\Lambda}\partial_t W,
     r^2e^{-\Phi}\partial_t V,
     r^2e^{-\Phi}\partial_t{\cal Y}\sin\theta\right), 
     \label{eq:du_i} 
\end{eqnarray}
from which one can show that
\begin{align}
  \delta u^{\alpha}_{\ ;\alpha} 
   &= re^{-\Phi} \partial_{tr}W
      + re^{-\Phi}\left(\Lambda_{,r} + \frac{3}{r}\right)\partial_t W
      + e^{-\Phi}\partial_{t\theta} V 
      + \frac{\cos\theta}{\sin\theta}e^{-\Phi}\partial_t V.
\end{align}
%%%%%%%%
We note that $\delta u^{\alpha}_{\ ;\alpha}$ is composed of only polar perturbations and $\delta u^\alpha_{\ ;\alpha}=0$ even for non-axisymmetric axial perturbations.

Due to the spherically-symmetric background, linear perturbations with axial parity (torsional $t$- oscillations) are completely decoupled from those with polar parity (interface $i$-, shear $s$-, fundamental $f$-, and pressure $p$-modes). However, if one considers higher-order perturbations,  the axial and polar sectors inevitably couple via non-linear effects. In this study, as a first step, we examine the simplest case with non-linear effects, i.e., 2nd-order perturbations for the torsional oscillations induced by the coupling between the linear perturbations of both axial and polar parity.
We note that one can similarly consider the 2nd-order polar perturbations, which are also induced by linear polar and axial perturbations. But, in this study, we try to understand the magnetar QPOs via crustal oscillations (or at least subspectra due to such components), where most effort has gone into axial modes (again cf. crust-core couplings and global modes \cite{lev07a}). Meanwhile, the problem is a bit more involved for polar modes, so we leave it for future work (along with magnetic fields etc), and we focus on 2nd-order axial perturbations in this study.

The total energy-momentum tensor, $T_{\mu\nu}$, is composed of two parts: a fluid part, $T_{\mu\nu}^{\rm (F)}$, and a shear part, $T_{\mu\nu}^{\rm (S)}$. One can derive the fluid part for the $\phi$-component of the perturbed energy-momentum conservation law, which contributes to the torsional oscillations, as
\begin{align}
  \delta T^{{\rm (F)}\phi\nu}_{\ \ \ \ \ \ ;\nu} 
     =& \frac{(\varepsilon + p)e^{-2\Phi}}{\sin\theta}
        \bigg\{\left[1+ \frac{\delta \varepsilon + \delta p}{\varepsilon + p}\right]\partial_{tt}{\cal Y}
     + r(\partial_t W)(\partial_{tr}{\cal Y})
     + (\partial_t V)(\partial_{t\theta}{\cal Y})
     + \left(r\partial_{tr} W+\partial_{t\theta} V\right)(\partial_t{\cal Y}) \nonumber \\
     &+\frac{\partial_t(\delta \varepsilon + \delta p)}{\varepsilon + p}\partial_t {\cal Y}
     + \left[\frac{\varepsilon' + p'}{\varepsilon + p} - \Phi_{,r}  + \Lambda_{,r} 
     + \frac{5}{r}\right]r(\partial_t W)(\partial_t{\cal Y})
     + \frac{2\cos\theta}{\sin\theta}(\partial_t V)(\partial_{t}{\cal Y})\bigg\} \,,
\end{align}
where we omit third- and higher-order terms, such as $\delta \varepsilon(\partial_t W)(\partial_t{\cal Y}) \sim \mathcal{O}(\xi^3)$.

Regarding the contribution from the shear stress, we simply follow the Schumaker-Thorne formalism \cite{Schumaker83}, which employs the Hookean approximation so that the shear contribution to the stress-energy tensor from the shear tensor, $S_{\mu\nu}$, is proportional only to the shear modulus, i.e.,
%%%%%%
\begin{equation}
  T_{\mu\nu}^{\rm (S)} = -2\mu S_{\mu\nu}. \label{eq:Hook}
\end{equation}
%%%%%%
The shear tensor is related to the ``rate of shear'' tensor, $\sigma_{\mu\nu}$, \cite{CQ72} 
%%%%%%$\sigma_{\mu\nu}$, via~\cite{CQ72} 
\begin{equation}
  {\cal L}_{u}S_{\mu\nu} = \frac{2}{3}S_{\mu\nu}\nabla_\alpha u^\alpha 
     + \sigma_{\mu\nu},  \label{eq:Shear}
\end{equation}
%%%%%%
where ${\cal L}_u$ denotes the Lie derivative along the direction of four-velocity, $u^\mu$, and one has
%%%%%%
\begin{eqnarray}
  {\cal L}_u S_{\mu\nu} &=& u^\alpha \nabla_\alpha S_{\mu\nu} 
     + S_{\alpha\nu}\nabla_\mu u^{\alpha} + S_{\mu\alpha} \nabla_{\nu} u^{\alpha}, \\
  \sigma_{\mu\nu} &=& \frac{1}{2}\left(P^\alpha_{\ \nu}\nabla_\alpha u_\mu
     + P^\alpha_{\ \mu}\nabla_\alpha u_\nu\right) 
     - \frac{1}{3}P_{\mu\nu}\nabla_\alpha u^\alpha \, . \label{eq:sigma}
\end{eqnarray}
%%%%%%
Here, $P_{\mu\nu}$ is the projection tensor given by
%%%%%%
\begin{equation}
  P_{\mu\nu} = g_{\mu\nu} + u_\mu u_\nu \, . \label{eq:projection}
\end{equation}
%%%%%%
%%
From Eq. (\ref{eq:Shear}), one can get 
%%%%%%%%%
\begin{equation}
  \delta({\cal L}_uS_{\mu\nu}) = \delta \sigma_{\mu\nu} + \frac{2}{3}\delta S_{\mu\nu}\delta u^\alpha_{\ ;\alpha}, \label{eq:Shear1}
\end{equation}
%%%%%%%%
because $u^\alpha_{\ ;\alpha}=0$ and $S_{\mu\nu}=0$ for the (background) equilibrium model. Running through the components $(\mu,\nu)=(\mu,\phi)$ in Eq.~(\ref{eq:Shear1}), it is possible to find
%%%%%%%%
\begin{align}
  \delta \sigma_{t\phi} + \frac{2}{3}\delta S_{t\phi}\delta u^\alpha_{\ ;\alpha}
    =& e^{-\Phi} \delta S_{t\phi,t} 
        + \delta u^r \delta S_{t\phi,r} 
        + \delta u^\theta \delta S_{t\phi,\theta} 
        + \delta S_{r\phi} \delta u^r_{\ ,t} 
        + \delta S_{\theta\phi} \delta u^\theta_{\ ,t} 
        + \delta S_{\phi\phi} \delta u^\phi_{\ ,t},  \label{eq:S03-s03}  \\
  \delta \sigma_{r\phi} + \frac{2}{3}\delta S_{r\phi}\delta u^\alpha_{\ ;\alpha}
    =& e^{-\Phi} (\delta S_{r\phi,t} - \Phi_{,r} \delta S_{t\phi})
        + \delta u^r \delta S_{r\phi,r}
        + \delta u^\theta \delta S_{r\phi,\theta} 
        + \delta S_{r\phi} \delta u^r_{\ ,r}
        + \delta S_{\theta\phi} \delta u^\theta_{\ ,r}
        + \delta S_{\phi\phi} \delta u^\phi_{\ ,r},  \label{eq:S13-s13} \\
  \delta \sigma_{\theta\phi} + \frac{2}{3}\delta S_{\theta\phi}\delta u^\alpha_{\ ;\alpha}
    =& e^{-\Phi} \delta S_{\theta\phi,t} 
        + \delta u^r \delta S_{\theta\phi,r}
        + \delta u^\theta \delta S_{\theta\phi,\theta}
        + \delta S_{r\phi} \delta u^r_{\ ,\theta} 
        + \delta S_{\theta\phi} \delta u^\theta_{\ ,\theta}
        + \delta S_{\phi\phi} \delta u^\phi_{\ ,\theta},   \label{eq:S23-s23}  \\
  \delta \sigma_{\phi\phi} + \frac{2}{3}\delta S_{\phi\phi}\delta u^\alpha_{\ ;\alpha}
    =& e^{-\Phi} \delta S_{\phi\phi,t}
        + \delta u^r \delta S_{\phi\phi,r} 
        + \delta u^\theta \delta S_{\phi\phi,\theta},   \label{eq:S33-s33}
\end{align}
%%%%%%%%%
where $\delta \sigma_{\mu\phi}$ and $\delta u^{\alpha}_{\ ;\alpha}$ up to 2nd order in perturbations are given from the definitions as
%%%
\begin{align}
  \delta \sigma_{t\phi}
    =& \frac{e^\Phi}{2}\big[ 
       - \Phi_{,r} \delta u^r \delta u_\phi 
       - \delta u^r  \delta u_{\phi,r}
       - \delta u^\theta \delta u_{\phi,\theta}\big] 
       + \frac{1}{3} e^\Phi \delta u_\phi \delta u^\alpha_{\ ;\alpha}, \\
  \delta \sigma_{r\phi}
    =& \frac{1}{2}\left[\Phi_{,r} \delta u_\phi
       - \frac{2}{r}\delta u_\phi
       + \delta u_{\phi,r}
       + e^{-\Phi} \delta u_\phi\delta u_{r,t}
       + e^{-\Phi} \delta u_r \delta u_{\phi,t} \right],  \\
  \delta \sigma_{\theta\phi}
    =&  \frac{1}{2}\left[\delta u_{\phi,\theta}
       - 2\frac{\cos\theta}{\sin\theta} \delta u_\phi
       + e^{-\Phi}\delta u_\phi \delta u_{\theta,t}
       + e^{-\Phi} \delta u_\theta  \delta u_{\phi,t} \right], \\
  \delta \sigma_{\phi\phi}
    =& r e^{-2\Lambda}\sin^2\theta \delta u_r
       + \sin\theta\cos\theta \delta u_\theta 
       - \frac{1}{3}r^2\sin^2\theta \delta u^\alpha_{\ ;\alpha}
       + e^{-\Phi} \delta u_\phi \delta u_{\phi,t},   \\
  \delta u^{\alpha}_{\ ;\alpha}
   =& re^{-\Phi} \partial_{tr}W
      + re^{-\Phi}\left(\Lambda_{,r} + \frac{3}{r}\right)\partial_t W
      + e^{-\Phi}\partial_{t\theta} V 
      + \frac{\cos\theta}{\sin\theta}e^{-\Phi}\partial_t V. 
\end{align}
%%%

%%%%%%%%%%%%%%%%%%%%%%%%%%%%%%%%%
\subsection{Order matching}
%%%%%%%%%%%%%%%%%%%%%%%%%%%%%%%%%

At this stage, it is instructive to expand the relevant quantities up to successive orders. We recall that we consider only axial perturbations up to second-order, with the polar parity ones truncated at linear order. That is, $W^{(2)} = V^{(2)} = 0$ but ${\cal Y}^{(2)} \neq 0$. Some algebra now reveals that
%%%%%%%%%%%
\begin{align}
  \delta \sigma_{t\phi}^{(1)} =& 0\, , \\
  \delta \sigma_{r\phi}^{(1)} 
    =& \frac{r^2}{2}e^{-\Phi}\sin\theta(\partial_{tr}{\cal Y}^{(1)})\, ,   \\
  \delta \sigma_{\theta\phi}^{(1)}
    =& \frac{r^2}{2}e^{-\Phi}\left[\sin\theta(\partial_{t\theta}{\cal Y}^{(1)})
       -\cos\theta(\partial_t{\cal Y}^{(1)})\right] \, ,  \\
  \delta \sigma_{\phi\phi}^{(1)}
    =& -\frac{r^2}{3}e^{-\Phi}\sin^2\theta\left[r\partial_{tr}W^{(1)} + r\Lambda_{,r}\partial_t W^{(1)}
       + \partial_{t\theta}V^{(1)} - \frac{2\cos\theta}{\sin\theta}\partial_t V^{(1)}\right]\, ,    \\
  \delta u^{\alpha(1)}_{\ ;\alpha}
   =& re^{-\Phi} \partial_{tr}W^{(1)}
      + re^{-\Phi}\left(\Lambda_{,r} + \frac{3}{r}\right)\partial_t W^{(1)}
      + e^{-\Phi}\partial_{t\theta} V^{(1)}
      + \frac{\cos\theta}{\sin\theta}e^{-\Phi}\partial_t V^{(1)} \,,  \label{eq:duaa}
\end{align}
%%%%%%%%%%%
while
%%%%%%%%%%%
\begin{align}
  \delta \sigma_{t\phi}^{(2)}
    =& {\mathfrak s}_0({\cal P},{\cal A})\, ,  \\
  \delta \sigma_{r\phi}^{(2)}
    =& \frac{r^2}{2}e^{-\Phi}\sin\theta(\partial_{tr}{\cal Y}^{(2)}) 
       + {\mathfrak s}_1({\cal P},{\cal A})\, ,  \\
  \delta \sigma_{\theta\phi}^{(2)}
    =&  \frac{r^2}{2}e^{-\Phi}\left[\sin\theta(\partial_{t\theta}{\cal Y}^{(2)})
       -\cos\theta(\partial_t{\cal Y}^{(2)})\right]
       + {\mathfrak s}_2({\cal P},{\cal A})\, , \\
  \delta \sigma_{\phi\phi}^{(2)}
    =& r e^{-2\Lambda}\sin^2\theta \delta u_r^{(2)}
       + \sin\theta\cos\theta \delta u_\theta^{(2)} 
       - \frac{1}{3}r^2\sin^2\theta \delta u^{\alpha(2)}_{\ ;\alpha}
       + e^{-\Phi} \delta u_\phi^{(1)} \delta u_{\phi,t}^{(1)}\, ,
\end{align}
%%%%%%%%%%%
where ${\mathfrak s}_0({\cal P},{\cal A})$, ${\mathfrak s}_1({\cal P},{\cal A})$, and ${\mathfrak s}_2({\cal P},{\cal A})$ are the terms composed of the combination of the linear polar ($\mathcal{P} = V^{(1)}, W^{(1)}$) and linear axial perturbations ($\mathcal{A} = \mathcal{Y}^{(1)}$), whose forms are concretely given in Appendix~\ref{sec:appendix_1}. 
%We note that $\delta u^{\alpha(1)}_{\ ;\alpha}$ is only composed of linear polar perturbations. 

At linear-order, Eqs.~\eqref{eq:S03-s03} -- \eqref{eq:S33-s33} become
%%%%%%%%%%%%%%%%%%%%%%%%%%%%%%%%%%%%
%%%%%%%%%%%
\begin{eqnarray}
  \delta S^{(1)}_{t\phi}  &=& 0\, , \\
  \delta S^{(1)}_{r\phi}  &=& \frac{r^2}{2}\sin\theta (\partial_{r}{\cal Y}^{(1)})\, , 
      \label{eq:13a} \\
  \delta S^{(1)}_{\theta\phi}  
    &=&  \frac{r^2}{2}\left[\sin\theta(\partial_{\theta}{\cal Y}^{(1)}) - \cos\theta\, {\cal Y}^{(1)}\right]\, ,
       \label{eq:23a} \\
  \delta S^{(1)}_{\phi\phi}
    &=&  -\frac{r^2}{3}\sin^2\theta\left[r\partial_{r}W^{(1)} + r\Lambda_{,r} W^{(1)}
       + \partial_{\theta}V^{(1)} - \frac{2\cos\theta}{\sin\theta}V^{(1)}\right]. \label{eq:33a}
\end{eqnarray}
%%%%%%%%%%%%%%%%%%%%%%%%%%%%%%%%%%%%
We assume that $\delta S^{(1)}_{\mu\nu}= 0$ at $t=0$, i.e., stress builds up from zero from a relaxed background state. We note that $\delta S_{r\phi}^{(1)}$ and $\delta S_{\theta\phi}^{(1)}$ are only composed of linear axial perturbations, while $\delta S_{\phi\phi}^{(1)}$ is only composed of linear polar perturbations. Further note that the other components $\delta S^{(1)}_{ij}$ are non-zero, but do not contribute to the torsional-mode dynamics in a static, unmagnetized star; they can be found elsewhere in the literature (e.g. \cite{col12}).

Moving on to second-order, Eqs.~\eqref{eq:S03-s03} -- \eqref{eq:S23-s23} become
%%%%%%%%%%%
\begin{align}
  \delta S_{t\phi,t}^{(2)}
    =& e^{\Phi}\left[\delta \sigma_{t\phi}^{(2)}
        - \delta S_{r\phi}^{(1)} \delta u^{r(1)}_{\ ,t} 
        - \delta S_{\theta\phi}^{(1)} \delta u^{\theta(1)}_{\ ,t} 
        - \delta S_{\phi\phi}^{(1)} \delta u^{\phi(1)}_{\ ,t}\right],  \label{eq:S03b1}  \\
  \delta S_{r\phi,t}^{(2)} 
    =& \Phi_{,r} \delta S_{t\phi}^{(2)}
        +e^{\Phi}\left[ \delta \sigma_{r\phi}^{(2)} + \frac{2}{3}\delta S_{r\phi}^{(1)} \delta u^{\alpha(1)}_{\ ;\alpha}
        - \delta u^{r(1)} \delta S_{r\phi,r}^{(1)}
        - \delta u^{\theta(1)} \delta S_{r\phi,\theta}^{(1)} 
        - \delta S_{r\phi}^{(1)} \delta u^{r(1)}_{\ ,r}
        - \delta S_{\theta\phi}^{(1)} \delta u^{\theta(1)}_{\ ,r}
        - \delta S_{\phi\phi}^{(1)} \delta u^{\phi(1)}_{\ ,r}\right],  \label{eq:S13b1} \\
  \delta S_{\theta\phi,t}^{(2)} 
    =& e^{\Phi}\left[\delta \sigma_{\theta\phi}^{(2)} 
        + \frac{2}{3}\delta S_{\theta\phi}^{(1)} \delta u^{\alpha(1)}_{\ ;\alpha}
        - \delta u^{r(1)} \delta S_{\theta\phi,r}^{(1)}
        - \delta u^{\theta(1)} \delta S_{\theta\phi,\theta}^{(1)}
        - \delta S_{r\phi}^{(1)} \delta u^{r(1)}_{\ ,\theta} 
        - \delta S_{\theta\phi}^{(1)} \delta u^{\theta(1)}_{\ ,\theta}
        - \delta S_{\phi\phi}^{(1)} \delta u^{\phi(1)}_{\ ,\theta}\right].   \label{eq:S23b1} 
\end{align}
We note that the second-order perturbation, $\delta S_{\phi\phi}^{(2)}$, does not contribute to the second-order equations for torsional oscillations. From Eqs. (\ref{eq:S03b1}) -- (\ref{eq:S23b1}), one can derive  
\begin{align}
  \delta S_{t\phi,t}^{(2)}  \equiv& {\mathfrak S}_0({\cal P},{\cal A}),  \\
  \delta S_{r\phi,t}^{(2)} =& \frac{r^2}{2}\sin\theta(\partial_{tr}{\cal Y}^{(2)}) 
       + {\mathfrak S}_1({\cal P},{\cal A})
       + \Phi_{,r} \delta S_{t\phi}^{(2)}, \\
  \delta S_{\theta\phi,t}^{(2)} 
     =& \frac{r^2}{2}\left[\sin\theta(\partial_{t\theta}{\cal Y}^{(2)})
        -\cos\theta(\partial_t{\cal Y}^{(2)})\right]
        + {\mathfrak S}_2({\cal P},{\cal A}),
\end{align}
%%%%%%%%%%%
where again the Gothic symbols ${\mathfrak S}_0({\cal P},{\cal A})$, ${\mathfrak S}_1({\cal P},{\cal A})$, and ${\mathfrak S}_2({\cal P},{\cal A})$ represent terms composed of combinations of the linear polar and linear axial perturbations; they are shown explicitly in Appendix \ref{sec:appendix_1}. These equations can be immediately integrated, again imposing the initial condition that we start with an unstressed state at all orders. By defining ${{\cal S}}_i({\cal P},{\cal A})$ for $i=0,1,2$ as
\begin{align}
  {\cal S}_i({\cal P},{\cal A}) \equiv \int_0^t{\mathfrak S}_i({\cal P},{\cal A})dt, 
\end{align}
%%%%%%%%%%%
one can get
%%%%%%%%%%%
\begin{align}
  \delta S^{(2)}_{t\phi} =&  {{\cal S}}_0({\cal P},{\cal A})\, , \\
  \delta S^{(2)}_{r\phi} =&  \frac{r^2}{2}\sin\theta(\partial_{r}{\cal Y}^{(2)}) 
     + {\cal S}_1({\cal P},{\cal A}) + \Phi_{,r} \int_0^t \delta S_{t\phi}^{(2)} dt\, , \\
  \delta S^{(2)}_{\theta\phi} =& \frac{r^2}{2}\left[\sin\theta(\partial_{\theta}{\cal Y}^{(2)})
     -\cos\theta\, {\cal Y}^{(2)}\right] + {\cal S}_2({\cal P},{\cal A})\, .
\end{align}
%%%%%%%%%%%

Now, simply\footnote{In considering second-order perturbations of the shear portion of the energy-momentum tensor we could, in principle, have terms that are proportional to $\delta\mu$. Such terms could be self-consistently associated with the density perturbation, $\delta \varepsilon$, implicitly through Eq.~\eqref{eq:mu}. This is not trivial however since the shear modulus depends only on the \emph{ion density}, rather than the overall density, $\delta \varepsilon$. In practice, if one considers a contribution from $\delta\mu$, one has to add the terms $\delta\mu^{(1)}\delta S_{r\phi}^{(1)}$ to Eq. (\ref{eq:dTs13}) and $\delta\mu^{(1)}\delta S_{\theta\phi}^{(1)}$ to Eq. (\ref{eq:dTs23}). These additional terms change the source term ${\cal S}_S({\cal P},{\cal A})$ in Eq. (\ref{eq:2nda}) a little, but as mentioned below, the main results shown in this study are unchanged.} considering $\delta T^{\rm (S)}_{\mu\nu} = -2\mu\delta S_{\mu\nu}$
from Eq. (\ref{eq:Hook}), i.e., assuming that $\delta\mu=0$, we find  its linear components
%%%%%%%%%%%
\begin{align}
  \delta T^{{\rm (S)}(1)}_{t\phi} =&  0 \,, \\
  \delta T^{{\rm (S)}(1)}_{r\phi} 
     =&  -\mu r^2 \sin\theta (\partial_{r}{\cal Y}^{(1)}) \,, \\
  \delta T^{{\rm (S)}(1)}_{\theta\phi} 
     =&  -\mu r^2 \left[\sin\theta(\partial_{\theta}{\cal Y}^{(1)}) - \cos\theta\, {\cal Y}^{(1)}\right]\, , 
\end{align}
%%%%%%%%%%%
 as well as its second-order components 
%%%%%%%%%%%
\begin{align}
  \delta T^{{\rm (S)}(2)}_{t\phi} 
     =& -2\mu {\cal S}_0({\cal P},{\cal A})\, , \\
  \delta T^{{\rm (S)}(2)}_{r\phi} 
     =& -\mu r^2 \sin\theta(\partial_{r}{\cal Y}^{(2)}) - 2\mu {\cal S}_1({\cal P},{\cal A})
        - 2\mu \Phi_{,r}\int_0^t \delta S_{t\phi}^{(2)} dt, \label{eq:dTs13} \\
  \delta T^{{\rm (S)}(2)}_{\theta\phi}
     =&-\mu r^2 \left[\sin\theta(\partial_{\theta}{\cal Y}^{(2)})
        -\cos\theta\, {\cal Y}^{(2)}\right] - 2\mu {\cal S}_2({\cal P},{\cal A})\, .  \label{eq:dTs23}
\end{align}
%%%%%%%%%%%
The divergence of $\delta T^{{\rm (S)}(1)\phi\nu}$ is given as
%%%%%%%%%%%
\begin{align}
   \delta T^{{\rm (S)}(1)\phi\nu}_{\qquad \quad;\nu} 
       =& -\left[\frac{\mu e^{-2\Lambda}}{\sin\theta}(\partial_{r}{\cal Y}^{(1)})\right]_{,r}
         + \frac{\mu}{r^2}\left[\frac{\cos\theta}{\sin^2\theta}{\cal Y}^{(1)} 
            - \frac{1}{\sin\theta}(\partial_\theta{\cal Y}^{(1)})\right]_{,\theta} \nonumber \\
         &- \frac{\mu e^{-2\Lambda}}{\sin\theta}\left(\Phi_{,r}+\Lambda_{,r} + \frac{4}{r}\right)(\partial_r{\cal Y}^{(1)})
         + \frac{3\mu}{r^2}\left[\frac{\cos^2\theta}{\sin^3\theta}{\cal Y}^{(1)} 
            - \frac{\cos\theta}{\sin^2\theta}(\partial_\theta{\cal Y}^{(1)})\right]\, .
\end{align}
%%%%%%%%%%%
As is well-known, the relevant equations of motion are separable at first-order. Abusing notation slightly and decomposing ${\cal Y}^{(1)}(t,r,\theta)=\sum_{\ell} {\cal Y}_\ell^{(1)}(t,r)\partial_\theta P_\ell (\cos\theta)$ for Legendre polynomials $P_\ell(\cos\theta)$, one can express the divergence $\delta T^{{\rm (S)}(1)\phi\nu}_{\ \ \ \ \ \ \ \ ;\nu}$, for each $\ell$, as
%%%%%%%%%%%
\begin{align}
   \delta T^{{\rm (S)}(1)\phi\nu}_{\ \ \ \ \ \ \ \ ;\nu} 
       =& -\mu e^{-2\Lambda}\left[\partial_{rr}{\cal Y}_\ell^{(1)}
           + \left(\frac{\mu_{,r}}{\mu} + \Phi_{,r} - \Lambda_{,r} + \frac{4}{r}\right)(\partial_{r}{\cal Y}_\ell^{(1)}) 
           - \frac{(\ell+2)(\ell-1)}{r^2}e^{2\Lambda}{\cal Y}_\ell^{(1)}  \right]\frac{1}{\sin\theta}\partial_\theta P_\ell,
\end{align}
%%%%%%%%%%%
where we have used the well known relation $\left[\partial^2_\theta + (\cos\theta/\sin\theta) \partial_\theta + \ell(\ell+1)\right]P_\ell=0$. Thus, the perturbation equation at linear level reduces to 
%%%%%%%%%%%
\begin{align}
  0 =& \delta T^{{\rm (F)}(1)\phi\nu}_{\ \ \ \ \ \ \ \ ;\nu} + \delta T^{{\rm (S)}(1)\phi\nu}_{\ \ \ \ \ \ \ \ ;\nu} 
     \nonumber \\
   =& (\varepsilon + p)e^{-2\Phi}(\partial_{tt}{\cal Y}_\ell^{(1)})\frac{1}{\sin\theta}\partial_\theta P_\ell
     \nonumber \\
    &-\mu e^{-2\Lambda}\left[\partial_{rr}{\cal Y}_\ell^{(1)} 
           + \left(\frac{\mu_{,r}}{\mu} + \Phi_{,r} - \Lambda_{,r} + \frac{4}{r}\right)(\partial_{r}{\cal Y}_\ell^{(1)}) 
           - \frac{(\ell+2)(\ell-1)}{r^2}e^{2\Lambda}{\cal Y}_\ell^{(1)}  \right]\frac{1}{\sin\theta}\partial_\theta P_\ell, 
\end{align}
%%%%%%%%%%%
and hence 
%%%%%%%%%%%
\begin{align}
    0=-\frac{\varepsilon + p}{\mu}e^{-2\Phi+2\Lambda}(\partial_{tt}{\cal Y}_\ell^{(1)}) 
       + \partial_{rr}{\cal Y}_\ell^{(1)} 
       + \left(\frac{\mu_{,r}}{\mu} + \Phi_{,r} - \Lambda_{,r} + \frac{4}{r}\right)(\partial_{r}{\cal Y}_\ell^{(1)}) 
       - \frac{(\ell+2)(\ell-1)}{r^2}e^{2\Lambda}{\cal Y}_\ell^{(1)},  \label{eq:1st}
\end{align}
%%%%%%%%%%%
as in Ref.~\cite{Schumaker83}.

On the other hand, the divergence of $\delta T^{{\rm (S)}\phi\nu}$ at second-order reads
%%%%%%%%%%%
\begin{align}
   \delta T^{{\rm (S)}(2)\phi\nu}_{\ \ \ \ \ \ \ \ ;\nu} 
       =& \frac{\mu}{r^2\sin\theta}\bigg[
         - r^2e^{-2\Lambda} (\partial_{rr}{\cal Y}^{(2)})
         - r^2 e^{-2\Lambda} \left(\frac{\mu_{,r}}{\mu} +\Phi_{,r} - \Lambda_{,r} + \frac{4}{r}\right)
         (\partial_{r}{\cal Y}^{(2)}) \nonumber \\
         -& \partial_{\theta\theta}{\cal Y}^{(2)}
         -\cot\theta \partial_{\theta}{\cal Y}^{(2)}  
         +\left(\cot^2\theta-1\right) {\cal Y}^{(2)}  \bigg] 
         + {\cal S}_S({\cal P},{\cal A})\, ,  \label{eq:2nda}
\end{align}
%%%%%%%%%%%
where ${\cal S}_S({\cal P},{\cal A})$ is a coupling term composed of combinations of linear polar and axial perturbations, whose form is explicitly shown in Appendix~\ref{sec:appendix_1}.
Thus, the second-order perturbation equation for axial-parity oscillations can be derived from 
%%%%%%%%%%%
\begin{align}
  \delta T^{(2)\phi\nu}_{\qquad ;\nu} 
     = \delta T^{{\rm (F)}(2)\phi\nu}_{\qquad \, ;\nu} +  \delta T^{{\rm (S)}(2)\phi\nu}_{\qquad \, ;\nu} = 0, 
\end{align}
%%%%%%%%%%%
which leads to
%%%%%%%%%%%
\begin{align}
  -(\varepsilon + p)e^{-2\Phi+2\Lambda}(\partial_{tt}{\cal Y}^{(2)}) 
     +& \mu (\partial_{rr}{\cal Y}^{(2)})
     +\left[\mu_{,r} + \mu\left(\Phi_{,r} - \Lambda_{,r} + \frac{4}{r}\right)\right]
         (\partial_{r}{\cal Y}^{(2)}) \nonumber \\
     +& \frac{\mu}{r^2}e^{2\Lambda}\left[\partial_{\theta\theta}{\cal Y}^{(2)}
     +  \cot\theta \partial_{\theta}{\cal Y}^{(2)}   
     + \left(1 - \cot^2\theta\right) {\cal Y}^{(2)}\right]  
     = e^{2\Lambda}\sin\theta\left[{\cal S}_F({\cal P},{\cal A}) + {\cal S}_S({\cal P},{\cal A})\right],
     \label{eq:2nd-0}
\end{align}
%%%%%%%%%%%
where ${\cal S}_F({\cal P},{\cal A})$ is another part of the total coupling term composed of combinations of linear polar and axial perturbations, whose form is explicitly shown in Appendix~\ref{sec:appendix_1}.
Now, we introduce a new variable, $\tilde{\cal Y}(t,r,\theta)\equiv {\cal Y}(t,r,\theta)/\sin\theta$, which leads to $\tilde{\cal Y}^{(2)}= {\cal Y}^{(2)}/\sin\theta$. Then, Eq.~(\ref{eq:2nd-0}) becomes
%%%%%%%%%%%
 \begin{align}
  -(\varepsilon + p)e^{-2\Phi+2\Lambda}(\partial_{tt}\tilde{\cal Y}^{(2)})
     +& \mu (\partial_{rr}\tilde{\cal Y}^{(2)})
     +\left[\mu_{,r} + \mu\left(\Phi_{,r} - \Lambda_{,r} + \frac{4}{r}\right)\right]
         (\partial_{r}\tilde{\cal Y}^{(2)})        \nonumber \\
     +& \frac{\mu}{r^2}e^{2\Lambda}\left[(\partial_{\theta\theta}\tilde{\cal Y}^{(2)})
     +  3\partial_\theta\tilde{\cal Y}^{(2)}{\cot\theta}\right]  
     = e^{2\Lambda}\left[{\cal S}_F({\cal P},{\cal A}) + {\cal S}_S({\cal P},{\cal A})\right]. \label{eq:2nd}
\end{align}
%%%%%%%%%%%

We note that the operator defining Eq.~\eqref{eq:2nd} [i.e. the homogenous equation neglecting source terms] is identical to the linear equation for torsional oscillations. Once one selects specific linear oscillation modes for the axial and polar parity to be excited, one can examine the second-order spectrum. Furthermore, due to the coupling terms, Eq.~\eqref{eq:2nd} is generally not separable.

We close this section by stating that, in what remains of this paper, we numerically evolve Eq.~\eqref{eq:2nd-0} in isolation. That is, we do not solve a closed system where mode amplitudes adjust iteratively as seed energy is sapped from the first-order spectrum to grow modes at second-order as has been done, for example, in the $f$- \cite{pk15,pk16} and $r$-mode \cite{arr03,brink04} literatures. In reality, Eq.~\eqref{eq:1st} must be solved simultaneously to study the energy-transfer problem and onset of resonant or nonresonant parametric instabilities (i.e., parent-daughter-daughter couplings in addition to the direct parent-parent-daughter couplings). However, such complexity lies beyond the scope of this work, which is mostly concerned with deriving the relevant equations and illustrating how axial-polar couplings at first order can enrich the nonlinear spectrum.

%%%%%%%%%%%%%%%%%%%%%%%%%%%%%%%%%%%%%%%%%%%%%%%%
\subsection{Boundary and initial conditions}
\label{sec:BC_IC}
%%%%%%%%%%%%%%%%%%%%%%%%%%%%%%%%%%%%%%%%%%%%%%%%

The linear perturbation equation for torsional oscillations, Eq.~\eqref{eq:1st}, can be solved as an eigenvalue problem by taking a harmonic time-dependence $e^{i\omega t}$ for angular frequency $\omega = 2 \pi f$. We impose a zero traction condition at the base of the crust and the zero-torque condition at the interface between the crust and envelope (see e.g., Refs.~\cite{Schumaker83,SKS07} for details). On the other hand, the equations and boundary conditions for linear polar-parity oscillations are explicitly shown in Ref.~\cite{SL02,Sotani23}. With crustal elasticity, the interface ($i_i$-) and shear ($s_i$-) modes can be excited in addition to the usual acoustic oscillations, i.e., the fundamental ($f$-) and pressure ($p_i$-) modes. Since the number of excitable $i$-modes is generally the same as the number of interfaces, where the non-zero shear modulus discontinuously drops to zero, we can expect that two $i$-modes exist in our case. 
In addition, since we simply consider linear polar perturbations without stratification for zero-temperature nuclear matter in this study, we cannot discuss couplings induced by the gravity ($g$-) modes. However, high-overtones of $g$-modes are of low frequency, which may play an important role in explaining the observed QPOs in magnetar flares. Their inclusion is relatively straightforward, and will be investigated elsewhere.

Since it is not \emph{a priori} clear which multipoles $\ell$ will be excited by non-linear effects for some given initial data, we examine oscillations inside the crustal region over the whole angular domain $0\le \theta\le \pi$ in this study. 
To evolve Eq. (\ref{eq:2nd}) in two-dimensional space ($r$, $\theta$), one has to impose some boundary conditions. Here, as in the case of the linear perturbations with axial parity, we impose a zero traction condition at the base of the crust ($r=R_{c}$) and at the interface between the crust and envelope ($r=R_{e}$). In symbols, we enforce 
\begin{equation} \label{eq:bcr}
  \partial_r\tilde{\cal Y}^{(2)}|_{r=R_{c},R_{e}}=0,
\end{equation}
because of Eq.~(\ref{eq:dTs13}). On the other hand, since we simply consider axisymmetric perturbations, the boundary conditions imposed at $\theta = 0$ and $\pi$ should be $\partial_\theta \tilde{\cal Y}^{(2)}|_{\theta=0,\pi}=0$. The time evolution is calculated with the iterated Crank-Nicholson method \cite{iCN}. Hereafter, $N_r$ and $N_\theta$ denote the grid number in the radial and $\theta$ direction, respectively. 

In addition, we simply assume that ${\cal Y}^{(2)}=0$ at $t=0$ as an initial condition. That is, each ${\cal Y}^{(2)}$ solution shown below is completely induced by the source terms composed of the product of the linear axial and polar perturbations, with the coupling timescale being essentially instantaneous.

%%%%%%%%%%%%%%%%%%%%%%%%%%%%%%%%%%%%%%%%%%%%%%%%
\section{Results: Nonlinear Coupling}
\label{sec:2nd}
%%%%%%%%%%%%%%%%%%%%%%%%%%%%%%%%%%%%%%%%%%%%%%%%

The numerical results strongly depend on the spatial resolution. Numerical convergence tests, carried out by evolving Eq.~(\ref{eq:2nd}) assuming that ${\cal S}_F({\cal P},{\cal A})={\cal S}_S({\cal P},{\cal A})=0$ (i.e., the linear problem), are presented in Appendix \ref{sec:appendix_2}. These demonstrate how the results vary with radial, $N_r$, and angular, $N_\theta$, resolutions. Based on that analysis, we find that one adequately resolves the modes up to the 1st overtones with angular quantum-numbers reaching $\ell\simeq 10$ for $N_r \simeq 500$ and $N_\theta \simeq 100$. So, hereafter we adopt $N_r=500$ and $N_\theta = 120$ for evolutions with nonlinear coupling. We note that higher-order ($n>1$) overtones may not be resolved with our resolution even if they are excited, but these lie at far-too-high frequencies to be observable.  If considering problems with energy-transfer, these could still be relevant however (as for, e.g., $p$-$g$ couplings \cite{reed16}).

We now turn to the main results of this work and examine the torsional oscillations induced by non-linear couplings between linear perturbations of both axial and polar parity by evolving Eq.~(\ref{eq:2nd}) with the source terms, ${\cal S}_F({\cal P},{\cal A})$ and ${\cal S}_S({\cal P},{\cal A})$. As mentioned above, order-$\ell$ linear perturbations are independent of all other order-$\ell'\neq\ell$ linear perturbations. The polar perturbations are also independent of the axial ones. So, one can examine the pattern of mode excitation induced by the linear coupling via Eq.~(\ref{eq:2nd}) by selecting a specific combination of axial and polar oscillation modes, e.g., ${}_2t_0$ (the $\ell=2$ fundamental torsional mode) and ${}_2s_1$ (the $\ell=2$ first shear mode). In this study, as a first step, we consider the coupling between the torsional oscillations, ${}_\ell t_0$ (or ${}_\ell t_1$), and the $\ell=2$ polar oscillations (though we could consider modes of multipolarity even higher than $\ell = 11$ for $N_{\theta} = 120$, as detailed in Appendix \ref{sec:appendix_2}). A study involving other couplings may be shown elsewhere.

Several eigenfrequencies determined via the eigenvalue problem with a linear perturbation analysis, using the neutron star model constructed in Sec.~\ref{sec:background}, are listed in Table~\ref{tab:modes}.
We note that in this study we identify the $i_1$- and $i_2$-modes in such a way that the $i_2$-mode is the mode excited at the crust-envelope interface, while the $i_1$-mode (tangential) eigenfunction has jumps at both the crust-core and crust-envelope interfaces.
With the source terms, ${\cal S}_F({\cal P},{\cal A})$ and ${\cal S}_S({\cal P},{\cal A})$, determined explicitly using the quoted eigenfrequencies and associated eigenfunctions, we evolve Eq.~\eqref{eq:2nd} for 10 seconds.
For this, we expect a bandwidth of $\sim 0.1~$Hz. 
Even if we select a specific combination of linear axial and polar oscillation modes, there is still freedom in setting their amplitudes. In this study, we simply set the maximum amplitude of the selected linear axial (${\cal Y}^{(1)}$) and polar perturbations (either $W^{(1)}$ or $V^{(1)}$) in the elastic region to be unity. On the other hand, since the source terms, ${\cal S}_F({\cal P},{\cal A})$ and ${\cal S}_S({\cal P},{\cal A})$, are always composed of the product of the amplitudes of linear axial and polar perturbations, it is not the individual amplitudes that are important but rather their product in the evolution of ${\cal Y}^{(2)}$. One can expect the results shown in this study would be qualitatively unchanged, even if one changes such a product into some value less than 1.

%%%%%%%%%%%%%%%%%%%%%%%%%%%%%%%%%%%
% Table 1
%%%%%%%%%%%%%%%%%%%%%%%%%%%%%%%%%%%
\begin{table}
\centering
\caption{Eigenfrequencies determined via the (first-order) eigenvalue problem for the neutron star model considered in this study, i.e., $M=1.41M_\odot$ and $R=11.68$~km constructed with the SLy4 EOS. The labelling of these modes is defined in Sec.~\ref{sec:perturbations}.
}
\begin{tabular}{c|c}
\hline\hline
  Mode  & Frequency (Hz)   \\
\hline
  ${}_2t_0$ &   23.9  \\  
  ${}_3t_0$ &   37.8  \\  
  ${}_4t_0$ &   50.7  \\  
  ${}_5t_0$ &   63.2  \\  
  ${}_6t_0$ &   75.6  \\  
\hline
  ${}_2t_1$ &   898.5  \\  
  ${}_3t_1$ &   899.0  \\  
  ${}_4t_1$ &   899.7  \\  
  ${}_5t_1$ &   900.6  \\  
  ${}_6t_1$ &   901.7  \\  
\hline
  ${}_2i_2$ &   40.9 \\  
  ${}_2i_1$ &   45.2 \\  
\hline
  ${}_2s_1$ &  895.1  \\  
  ${}_2s_2$ &  1463.8  \\  
  ${}_2s_3$ &  1806.5  \\  
  ${}_2s_4$ &  2276.9  \\  
  ${}_2s_5$ &  2829.0  \\  
\hline
  ${}_2f$ &  2407.1  \\  
\hline\hline
\end{tabular}
\label{tab:modes}
\end{table}
%%%%%%%%%%%%%%%%%%%%%%%%%%%%%%%%%%%

%%%%%%%%%%%%%%%%%%%%%%%%%%%%%%%%%%%
\subsection{Seeding via coupled lowest-$n$ modes}
%%%%%%%%%%%%%%%%%%%%%%%%%%%%%%%%%%%

\begin{figure}[tbp]
\begin{center}
\includegraphics[scale=0.5]{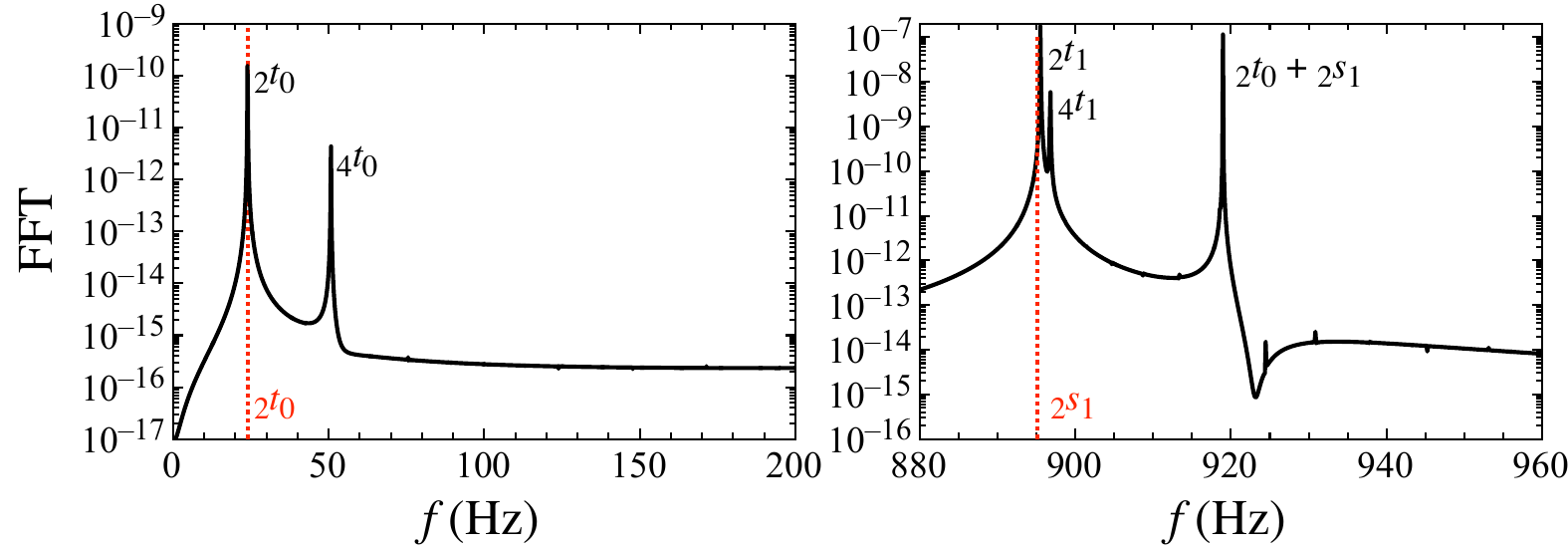}
\end{center}
\caption{%%
The FFT of $\mathcal{Y}^{(2)}$ evolved for 10 seconds. Here we imposed a coupling (i.e. initial data) between ${}_2t_0$ (23.9 Hz) and ${}_2s_1$ (895.1 Hz). For reference, the ${}_2t_0$ and ${}_2s_1$ frequencies are shown with dotted lines and we have focused on the regions $0 \leq f \leq 200$~Hz and $880 \leq f \le 960$~Hz as no modes are excited in the interval $200 \leq f \leq 880$~Hz.
}%%
\label{fig:FFT_2t0_2s1}
\end{figure}

First, we show the results with coupling between ${}_2t_0$ and ${}_2s_1$. The FFT computed via the resulting waveform for $\mathcal{Y}^{(2)}$ is shown in Fig.~\ref{fig:FFT_2t0_2s1}. We find several modes are excited, which we identify as ${}_2t_0$, ${}_4t_0$, ${}_2t_1$, and ${}_4t_1$. In addition to these frequencies, which appear in the linear perturbation analysis, we also find additional modes, whose frequencies correspond to the sum of that associated with ${}_2t_0$ and ${}_2s_1$. We note that the oscillations with ${}_2t_1$, ${}_4t_1$, and ${}_2t_0+{}_2s_1$ (= 919.0 Hz) seem to be excited more strongly than those with ${}_2t_0$ and ${}_4t_0$ at the 2nd-order perturbation level, even though we consider the ${}_2t_0$ mode to build the source terms. Note we are showing \emph{only} the second-order spectrum, not FFTs of the total eigenfunction $\mathcal{Y}$. More precisely, although we sometimes find $\mathcal{Y}^{(2)}_{\rm amp,daughter} > \mathcal{Y}^{(2)}_{\rm amp,parents}$, we still have $\mathcal{Y}^{(1)}_{\rm amp,daughter} + \mathcal{Y}^{(2)}_{\rm amp,daughter} = \mathcal{Y}^{(2)}_{\rm amp,daughter} \ll \mathcal{Y}^{(1)}_{\rm amp,parents} + \mathcal{Y}^{(2)}_{\rm amp,parents}$, where $\mathcal{Y}^{(i)}_{\rm amp,daughter}$ and $\mathcal{Y}^{(i)}_{\rm amp,parents}$ denote the amplitude of the $i$-th order perturbations for the daughter and parent modes, respectively.

%Old kept here; maybe it is better somehow a mixture between the two?: \upd{Even so, the amplitude of the total perturbations, i.e., ${\cal Y}={\cal Y}^{(1)} + {\cal Y}^{(2)}$, for the parent modes is much larger than those for the daughter modes, since the amplitude of the parent modes is on the order of unity in this study (and that of the daughter mode is 0). }

%%%%%%%%%%%%%%%%%%%%%%%%%%%%%%%%%%%
% Figure 1
%%%%%%%%%%%%%%%%%%%%%%%%%%%%%%%%%%%
%%%%%%%%%%%%%%%%%%%%%%%%%%%%%%%%%%%

%\sout{Although this may appear unexpected, \upd{e.g., why },} 
The reason why ${}_3t_0$ is not excited can be understood by carefully examining the perturbation equations \eqref{eq:2nd}. In general, the odd- and even-order Legendre polynomials for $\ell\ge 2$ can be written as
\begin{gather}
  P_{2m} = \sum_{j=0}^m a_{j}\cos^{2j}\theta, \\
  P_{2m+1} = \sum_{j=0}^m b_{j}\cos^{2j+1}\theta, 
\end{gather}
where $a_j$ and $b_j$ are appropriate constants. Meanwhile, the angular profile of polar and axial perturbations are essentially expressed as $P_{\ell_p}$ and $\partial_\theta P_{\ell_a} / \sin\theta$, respectively, where the indices $\ell$ for the polar and axial perturbations are explicitly shown as $\ell_p$ and $\ell_a$.  So, the coupling between the linear polar and linear axial perturbations in the source terms, ${\cal S}_{i=F,S}$ in Eq.~(\ref{eq:2nd}) for a single mode behave like 
%%%%%%%%%%%%%%
\begin{align}
  {\cal S}_i({\cal P},{\cal A}) &\simeq P_{\ell_p} \frac{\partial_\theta P_{\ell_a}}{\sin\theta}   
        \label{eq:SPA_0}  \\
     &\simeq \sum_{j=0}^{m_p+m_a}\alpha_j\left(\frac{\partial_\theta P_{2j}}{\sin\theta}\right) \ \ \ 
        {\rm for}\ (\ell_p,\ell_a)=(2m_p,2m_a),  \label{eq:SPA_1}  \\         
     &\simeq \sum_{j=0}^{m_p+m_a}\alpha_j\left(\frac{\partial_\theta P_{2j+1}}{\sin\theta}\right) \ \ \ 
        {\rm for}\ (\ell_p,\ell_a)=(2m_p,2m_a+1), \label{eq:SPA_2} \\         
     &\simeq \sum_{j=0}^{m_p+m_a}\alpha_j\left(\frac{\partial_\theta P_{2j+1}}{\sin\theta}\right) \ \ \ 
        {\rm for}\ (\ell_p,\ell_a)=(2m_p+1,2m_a),  \label{eq:SPA_3} \\         
     &\simeq \sum_{j=0}^{m_p+m_a+1}\alpha_j\left(\frac{\partial_\theta P_{2j}}{\sin\theta}\right) \ \ \ 
        {\rm for}\ (\ell_p,\ell_a)=(2m_p+1,2m_a+1), \label{eq:SPA_4} 
\end{align}
%%%%%%%%%%%%%%
where $\alpha_j$ is again some constant which is a combination of $a_j$ and $b_j$. That is, the source terms should induce axial-type oscillations up to order $\ell_p + \ell_a$. Moreover, from the above expression, we find that torsional oscillations with only even (odd) values of $\ell$ are selectively excited when one considers polar and axial couplings such that $\ell_p$ and $\ell_a$ are both even or both odd numbers (even and odd numbers or odd and even numbers). For example, only the $\ell=2$, 4, and 6 torsional oscillations can be excited at second-order by the coupling between $\ell_p=2$ and $\ell_a=4$ at first order; similarly, only the $\ell=3$ and 5 second-order torsional oscillations can be excited by the coupling between $\ell_p=2$ and $\ell_a=3$. %\textcolor{blue}{As for why the amplitude of the daughter mode (${}_2t_0+{}_2s_1$) becomes larger than those of the parents, this is likely an artifact due to our chose of unit amplitude (see Sec.~\ref{sec:2nd}). Without energy transfer, the amplitude of the daughter should effectively be set as the square of the parent amplitudes multiplied by an integral which measures the degree of overlap between the respective eigenfunctions (similar to that used in studies of tidally- or otherwise forced modes \cite{pk15,kuan21}). Thus, even if the overlap multiplier exceeds unity, the daughter mode amplitude should be less than that of the parent for realistic initial amplitudes $\ll 1$.}

%\upd{On the other hand, we could not understand the results that the amplitude of the daughter mode (${}_2t_0+{}_2s_1$) becomes larger than those of the parents in the second order level.}

In addition to these mode excitations, since the source term in Eq.~(\ref{eq:2nd}) is composed of a combination of linear polar and linear axial perturbations, which respectively depend on time as $\exp(i\omega_p t)$ and $\exp(i\omega_a t)$ with eigenfrequencies $\omega_p$ and $\omega_a$ of the considered linear perturbations, one can expect an additional mode with frequency $\omega=\omega_p+\omega_a$ will be excited. 
This is simply a property of inhomogeneous wave equations with harmonic source terms. Note that, in a full treatment of the non-linear problem, one may expect the negative branch, $\omega=|\omega_p-\omega_a|$, to be excited also. This is because the eigenvalue equations depend on $\omega^2$, and hence both positive and negative roots are valid solutions. If we were to include source terms for total modes including conjugates, we would excite all branches $(\omega = |\omega_p\pm\omega_a|)$, though we set the amplitudes of the conjugates to zero for simplicity.

%%%%%%%%%%%%%%%%%%%%%%%%%%%%%%%%%%%
% Figure 2
%%%%%%%%%%%%%%%%%%%%%%%%%%%%%%%%%%%
\begin{figure}[tbp]
\begin{center}
\includegraphics[scale=0.5]{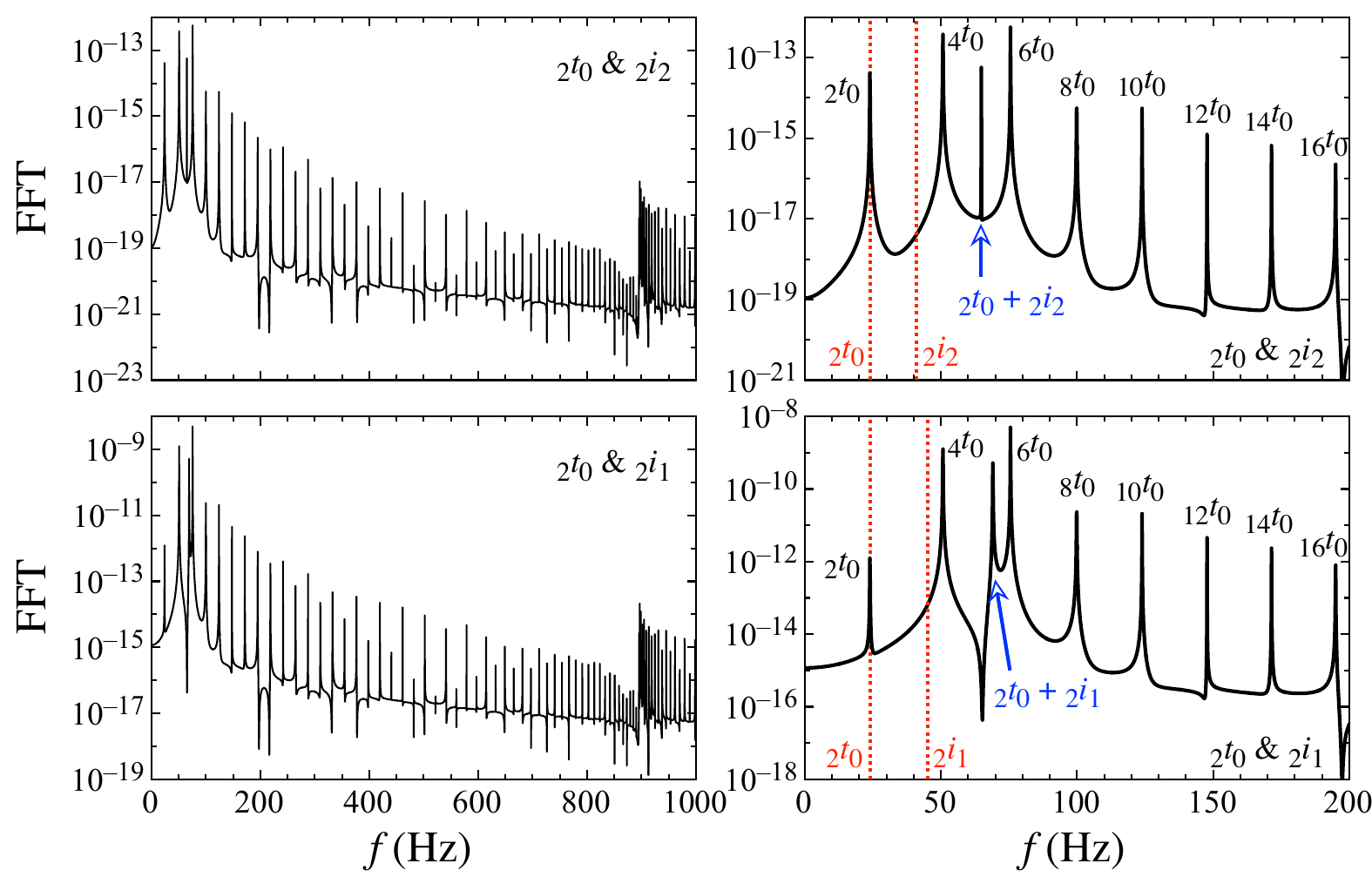}
\end{center}
\caption{%%
FFT from the waveform obtained with the coupling between ${}_2t_0$ (23.9 Hz) and ${}_2i_2$ (40.9 Hz) in the top panel, and between ${}_2t_0$ and ${}_2i_1$ (45.2 Hz) in the bottom panel. The right panels are just enlarged views of the left panels. For reference, in the right panel, we show the ${}_2t_0$, ${}_2i_2$, and ${}_2i_1$ frequencies with the dotted lines. 
}%%
\label{fig:FFT_2t0_2i}
\end{figure}
%%%%%%%%%%%%%%%%%%%%%%%%%%%%%%%%%%%

In a similar way, in Fig.~\ref{fig:FFT_2t0_2i}, we show the FFT obtained for an evolution triggered by the coupling between ${}_2t_0$ and ${}_2i_1$ oscillations (top panel), and between the ${}_2t_0$ and ${}_2i_2$ oscillations (bottom panel). As expected, we observe the additional mode excitation with a frequency of ${}_2t_0+{}_2i_1$ ($\approx$ 69.0 Hz) for the coupling between ${}_2t_0$ and ${}_2i_1$, and with a frequency of ${}_2t_0+{}_2i_2$ ($\approx$ 64.8 Hz) for the coupling between ${}_2t_0$ and ${}_2i_2$, highlighted by arrows in the right panels. However, in this case for the coupling with $i$-mode oscillations, we observe not only the $\ell=2$ and $\ell=4$ modes but also many other, even-order (in $\ell$) modes are significantly excited. This result cannot be immediately understood from the analytic arguments surrounding Eq.~\eqref{eq:SPA_1}. The richness of the excited spectrum may come from the fact that the frequency of the additionally ``superposed'' excitation mode, ${}_2t_0+{}_2i_1$ or ${}_2t_0+{}_2i_2$, is higher than ${}_4t_0$, and consequently the ${}_\ell t_0$-modes for $\ell\ge 6$ are also excited. 
In practice, in the coupling of the ${}_2t_0$ with the ${}_2i$-modes, the ${}_4t_0$ and ${}_6t_0$ modes, which are the two sides of the frequencies of ${}_2t_0+{}_2i_1$ or ${}_2t_0+{}_2i_2$, are strongly excited, compared to the other modes. 

\begin{figure}[tbp]
\begin{center}
\includegraphics[scale=0.5]{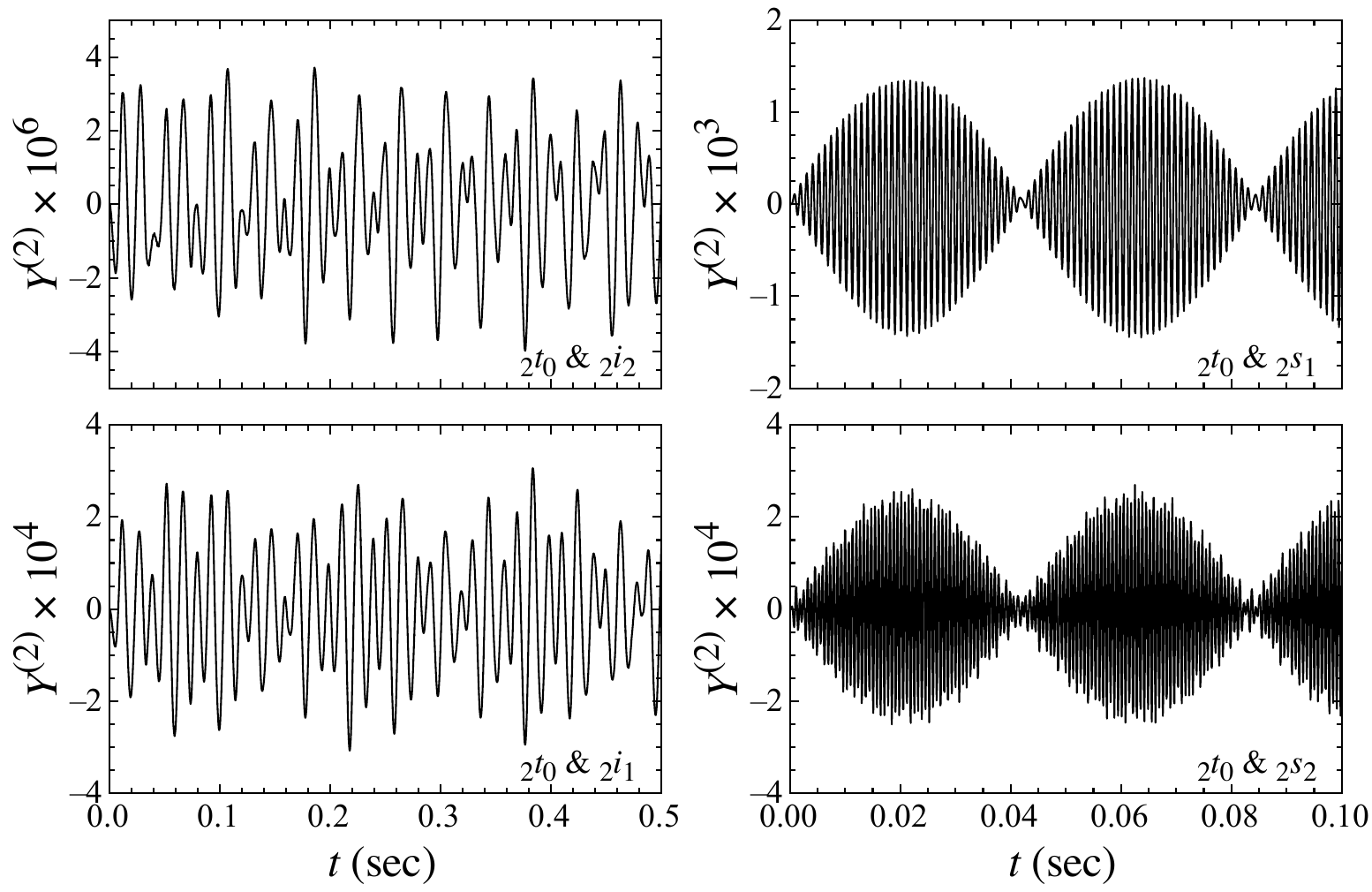}
\end{center}
\caption{%%
Waveforms of ${\cal Y}^{(2)}$, excited due to the coupling of ${}_2t_0$ with ${}_2i_2$ (top-left), ${}_2i_1$ (bottom-left), ${}_2s_1$ (top-right), or ${}_2s_2$ (bottom-right). We note that the duration shown here for the coupling with $i$-modes (0.5 sec) is different from that with $s$-modes (0.1 sec). 
}%%
\label{fig:wave-0t2}
\end{figure}

We also note that excitations at higher frequencies (overtones) have more power in the FFT for cases coupled to $s$-modes, while excitations with lower frequencies (fundamental oscillations) become stronger in cases with a coupling to an $i$-mode. 
Due to this feature, as shown in Fig.~\ref{fig:wave-0t2}, waveforms in cases with a linear-coupling to an $i$-mode (left panels) are of lower amplitude (note the scaling on the axes) from that for the coupling with $s$-mode (right panel).
We remind the reader that we have set the (initial) amplitudes as equal for the seeding modes, and these have magnitude~1.

%%%%%%%%%%%%%%%%%%%%%%%%%%%%%%%%%%%
% Figure 3
%%%%%%%%%%%%%%%%%%%%%%%%%%%%%%%%%%%
%%%%%%%%%%%%%%%%%%%%%%%%%%%%%%%%%%%

%%%%%%%%%%%%%%%%%%%%%%%%%%%%%%%%%%%
\subsection{Seeding via coupled overtones}
%%%%%%%%%%%%%%%%%%%%%%%%%%%%%%%%%%%

\begin{figure}[tbp]
\begin{center}
\includegraphics[scale=0.5]{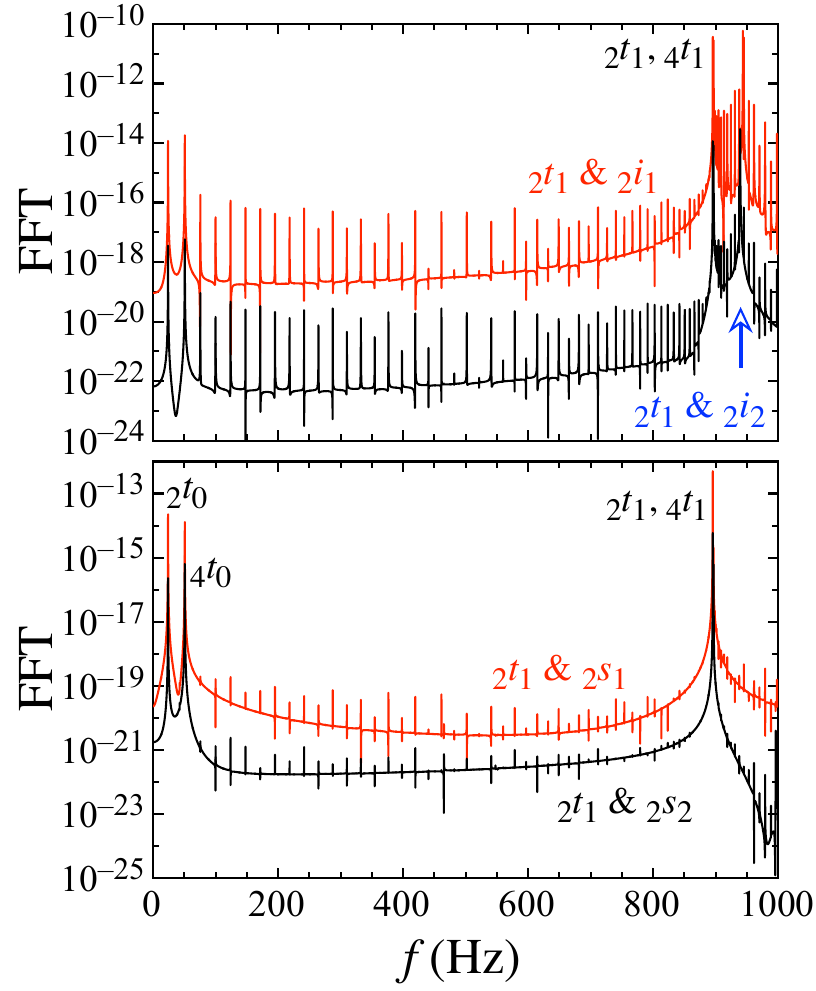}
\end{center}
\caption{%%
FFTs of ${\cal Y}^{(2)}$ obtained when considering the coupling of ${}_2t_1$ (898.5 Hz) with ${}_2i_2$ (40.9 Hz; black) or ${}_2i_1$ (45.2 Hz; red) in the top panel, or with ${}_2s_1$ (895.1 Hz; black) or ${}_2s_2$ (1463.8 Hz; red) in the bottom panel. In the top panel, the frequency denoted by an arrow corresponds to the frequency induced by the couplings, which do not appear in the linear analysis. 
}%%
\label{fig:2t1-2is}
\end{figure}

Next, in Fig.~\ref{fig:2t1-2is}, we show which modes are excited through the coupling of ${}_2 t_1$ (1st overtone of the torsional oscillations, rather than the fundamental mode) with an $i$-mode [${}_2i_2$ (black) or ${}_2i_1$ (red)] in the top panel and with an $s$-mode [${}_2s_1$ (black) or ${}_2s_2$ (red)] in the bottom panel, focusing on the frequency range lower than 1 kHz. 

Focusing first on the upper panel, we see that there are two closeby peaks at $\sim 900$~Hz. One of these corresponds to the ${}_2t_1$ and ${}_4t_1$ oscillations ($\approx 900$~Hz), while the additional frequency denoted with an arrow ($\approx 950$~Hz), which does not appear in the linear analysis, corresponds to ${}_2 t_1+{}_2i_2$ (${}_2 t_1+{}_2i_1$) for the coupling between ${}_2 t_1$ and ${}_2i_2$ (${}_2 t_1$ and ${}_2i_1$). The former of these persists in the case with a shear mode coupling, while the latter does not. We find that, unlike the case of the coupling of the fundamental torsional oscillations (Fig.~\ref{fig:FFT_2t0_2i}) with the $i$-modes, the torsional-mode overtones actually become stronger than the fundamental torsional modes, 
assuming that the amplitudes of the linear axial and polar perturbations in the source terms are equal to one. This is interesting astrophysically as it suggests that higher-frequency modes might get excited to larger amplitudes when considering higher-frequency seeds. In addition, since the frequency of the additional excitation (${}_2 t_1+{}_2i_2$ or ${}_2 t_1+{}_2i_1$) becomes much higher than those of the fundamental torsional oscillations, the strongest signals in the FFT among the fundamental modes are associated with ${}_2t_0$ and ${}_4t_0$. The feature in the coupling of ${}_2 t_1$ with the $s$-modes is almost the same as in the case of the coupling of ${}_2 t_0$, even though the frequency of the additional excitation lies outside of the frequency range considered in the figure.

%%%%%%%%%%%%%%%%%%%%%%%%%%%%%%%%%%%
% Figure 4
%%%%%%%%%%%%%%%%%%%%%%%%%%%%%%%%%%%
%%%%%%%%%%%%%%%%%%%%%%%%%%%%%%%%%%%

\begin{figure*}[tbp]
\begin{center}
\includegraphics[scale=0.35]{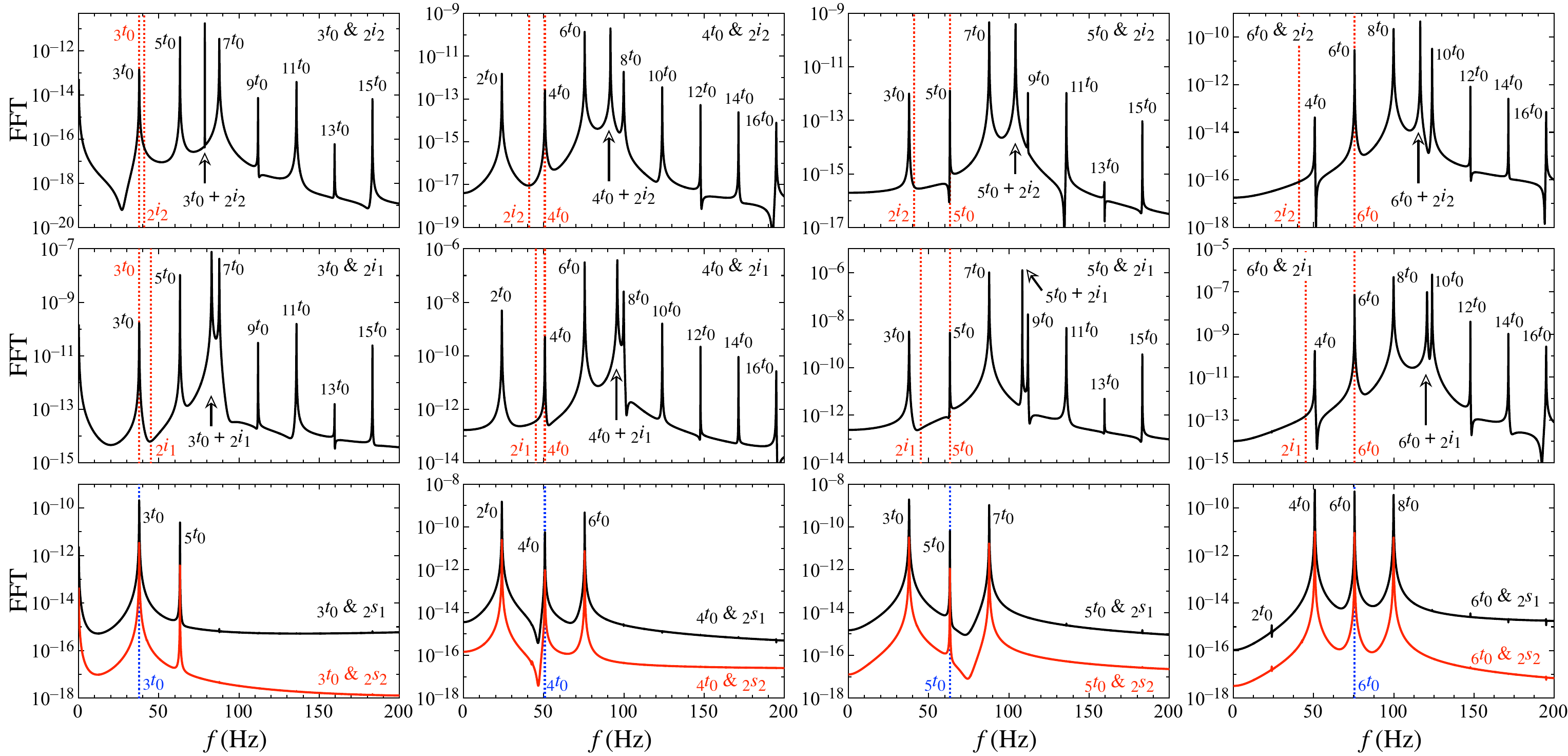}
\end{center}
\caption{%%
FFT from the waveform obtained with the coupling of ${}_\ell t_0$ for $\ell=3$, 4, 5, 6 with ${}_2i_2$ in the top panel, with ${}_2i_1$ in the middle panel, and with ${}_2s_i$ for $i=1,2$ in the bottom panel.
}%%
\label{fig:2i-3456t0}
\end{figure*}

To consider such phenomena more generally, we show in Fig.~\ref{fig:2i-3456t0} the FFTs computed from the waveforms of ${\cal Y}^{(2)}$ when considering seedings from the coupling of ${}_\ell t_0$ for $\ell=3-6$ with the ${}_2i_2$- (top panel), ${}_2i_2$- (middle panel), and ${}_2s_i$-modes for $i=1,2$ (bottom panel), focusing on the frequency range lower than 200 Hz to best illustrate the spectral structure. One observes a lot of mode excitations as well as the additional mode described earlier (with a frequency of ${}_\ell t_0+{}_2i_i$ for $i=1,2$ in the case of a coupling between ${}_\ell t_0$ and ${}_2i$-modes -- couplings with the shear modes are higher frequency). The strongest peaks correspond to this additional mode and those nearest neighbours in terms of frequency. In particular, the ${}_2t_0$-mode is not excited for the case of the coupling between ${}_6t_0$ and ${}_2 i$-modes,
since the ${}_2t_0$-mode frequency is located too far from the additional mode with the frequency of ${}_6 t_0+{}_2i_i$ for $i=1,2$. On the other hand, for the case of the coupling between the ${}_\ell t_0$ and ${}_2s$-modes, one can observe the excitations of the ${}_{\ell'} t_0$-modes with up to $\ell'=\ell+2$, as discuss with Eq.~(\ref{eq:SPA_0}). Even so, we find that three ${}_{\ell'} t_0$-modes with $\ell'=\ell-2$, $\ell$, and $\ell+2$ become much stronger signals for $\ell\ge 4$.

%%%%%%%%%%%%%%%%%%%%%%%%%%%%%%%%%%%
% Figure 5
%%%%%%%%%%%%%%%%%%%%%%%%%%%%%%%%%%%
%%%%%%%%%%%%%%%%%%%%%%%%%%%%%%%%%%%

\begin{figure*}[tbp]
\begin{center}
\includegraphics[scale=0.35]{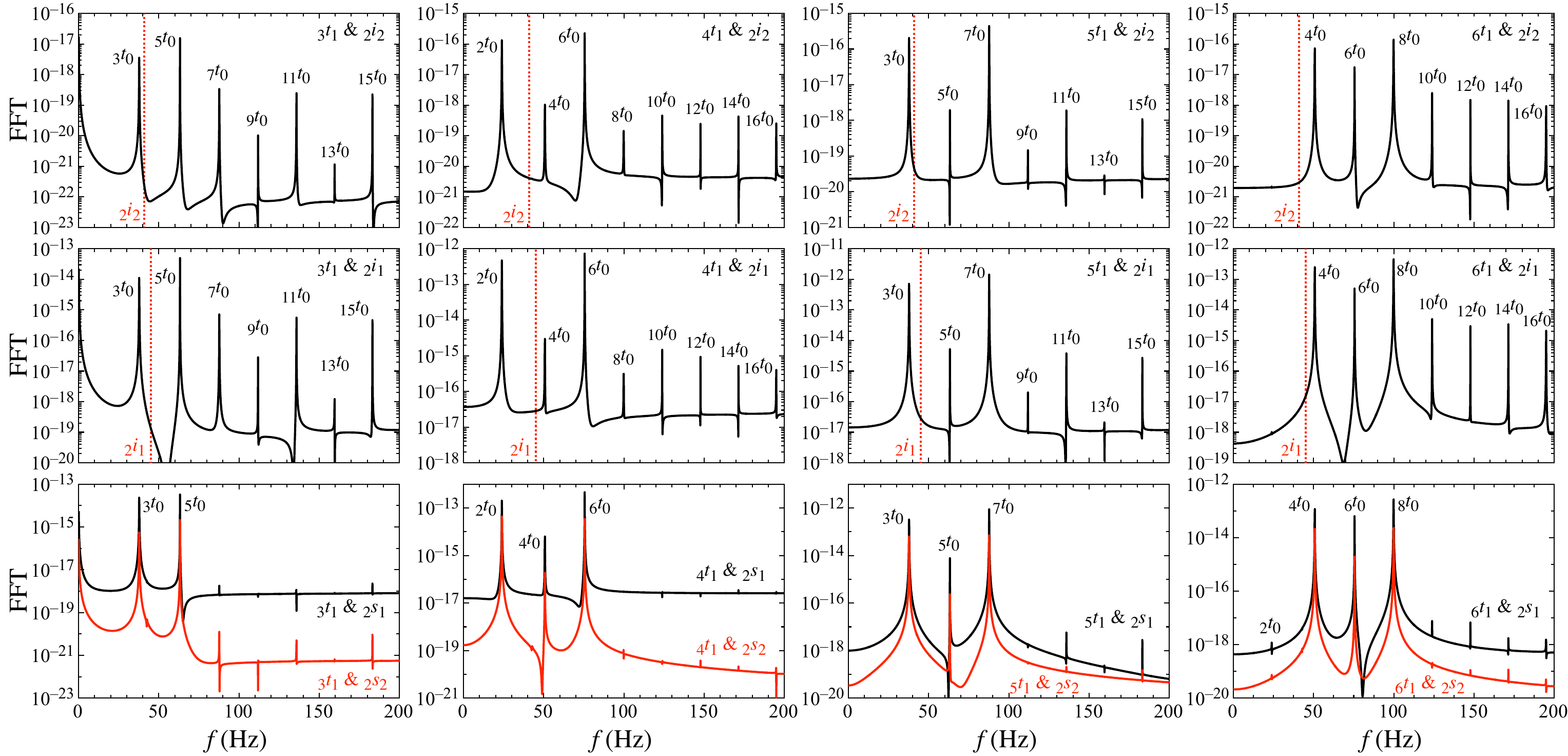}
\end{center}
\caption{%%
Same as Fig.~\ref{fig:2i-3456t0}, but showing FFTs for cases sourced by couplings with ${}_\ell t_1$ instead of ${}_\ell t_0$.
}%%
\label{fig:2i-3456t1}
\end{figure*}

On the other hand, Fig.~\ref{fig:2i-3456t1} is the same as Fig.~\ref{fig:2i-3456t0}, but with the coupling of ${}_\ell t_1$ instead of ${}_\ell t_0$ for $\ell=3$, 4, 5 and 6. In this figure, we focus only on the frequency range lower than 200 Hz, but we find that the overtones (and the additional excitation, which becomes in the frequency range of overtone now) are stronger than the fundamental oscillations, as shown in Fig.~\ref{fig:2t1-2is}. In the coupling of ${}_3 t_1$ with the ${}_2 i$-modes, we find that the modes of ${}_3 t_0$ and ${}_5 t_0$ are strongly excited as expected with Eq.~(\ref{eq:SPA_0}), unlike the case for the coupling with the fundamental torsional oscillations. In addition, for $\ell\ge4$ we find that the modes of ${}_{\ell'}t_0$ with $\ell'=\ell\pm 2$ become strong signals, while the mode with ${}_{\ell}t_0$ becomes weak. Meanwhile, in the coupling of ${}_\ell t_1$ with the ${}_2 s$-modes, the feature is more or less similar to the case for the coupling of ${}_\ell t_0$ as shown in the bottom panel of Fig.~\ref{fig:2i-3456t0}. We note that the frequency of ${}_\ell t_1$ weakly depends on the value of $\ell$, but which modes are excited due to the mode coupling discussed in this study strongly depends on the value of $\ell$.

%%%%%%%%%%%%%%%%%%%%%%%%%%%%%%%%%%%
% Figure 6
%%%%%%%%%%%%%%%%%%%%%%%%%%%%%%%%%%%
%%%%%%%%%%%%%%%%%%%%%%%%%%%%%%%%%%%

%%%%%%%%%%%%%%%%%%%%%%%%%%%%%%%%%%%%%%%%%%%%%%%%
\section{Conclusions}
\label{sec:Conclusion}
%%%%%%%%%%%%%%%%%%%%%%%%%%%%%%%%%%%%%%%%%%%%%%%%

In this paper, we have derived equations describing the evolution of second-order, axisymmetric torsional oscillations inside the elastic crust of a neutron star in general relativity. The relevant master equation, Eq.~(\ref{eq:2nd-0}), was solved in the time domain to determine the second-order displacement, ${\cal Y}^{(2)}$, for a variety of astrophysically-motivated initial conditions. Though we only evolve the second-order functions in the axial sector (i.e. we set $V^{(2)} = W^{(2)} = 0$), we self-consistently allow ${\cal Y}^{(2)}$ to be sourced by coupled axial-polar oscillations at first order. This represents a useful step forward towards the difficult inverse problem of determining neutron star properties from oscillations observed in the tails of giant flares or other transients. Indeed, most previous studies have focussed on the linear problem, while those that have been carried out at a non-linear level have not included seeding by coupled axial-polar parents. Though we have not considered magnetic fields or crust-core coupling in this paper, already the nonlinear analysis in our relatively simple case reveals some interesting features that are likely to apply to a real, astrophysical star.

One of these features concerns the nature of excited daughters from a given pair of parents. In agreement with the analytic predictions based on Legendre orthogonality (see Eqs.~\ref{eq:SPA_0}--\ref{eq:SPA_4}), a variety of modes can be excited from mixed couplings. Figs.~\ref{fig:2i-3456t0} and \ref{fig:2i-3456t1} demonstrate a rich spectrum from just two seed modes of different multipolarity or tonality (torsional plus either interface or shear modes), the amplitudes of which vary by several orders of magnitude and do not necessarily decrease with increasing $\ell$ or $n$ as one might naively expect. This implies that, in a real neutron star system where an elastic overstraining occurs, it is critical to account for nonlinear dynamics. 

There are several directions that would be worth pursuing in order to improve on this work. One of these concerns is feedback in the coupling. In this study we have determined the linear oscillation spectrum via the eigenvalue problem, the solutions of which then act as fixed (though still time-dependent) source terms in the second-order equation. However, a more realistic approach would be to loop these into each other to study energy transfer, as in the Newtonian scheme devised by Dziembowski \cite{dz80} and others (e.g., \cite{pk15,pk16}). By including the energy transfer dynamics one can investigate the saturation amplitude of various excited modes (see also Ref.~\cite{gab09}). These amplitudes can then be compared to the Bayes factors in various statistical analyses for the dynamic spectra in various giant flare tails (e.g. \cite{mill19}) to approach the inverse problem. Other obvious directions are to include higher-order polar couplings, magnetic fields, and the core. Such studies will be carried out elsewhere.

%\newpage
%%%%%%%%%%%%%%%%%%%%%%%%%%%%%%%%%%%%%%%%%%%%%%%%
\acknowledgments
%%%%%%%%%%%%%%%%%%%%%%%%%%%%%%%%%%%%%%%%%%%%%%%%
This work is supported in part by Japan Society for the Promotion of Science (JSPS) KAKENHI Grant Numbers 
JP19KK0354,     % International (A) by Sotani
JP21H01088,     % Kiban(B) by Sotani
JP23K20848,     % Kiban(B) by Sotani
and JP24KF0090, % by Sotani & Kumar
by FY2023 RIKEN Incentive Research Project,
and by Pioneering Program of RIKEN for Evolution of Matter in the Universe (r-EMU).
This work has been supported by the HORIZON-MSCA-2022 project ``EinsteinWaves (101131233)''.
AGS is supported by the Prometeo Excellence Programme grant CIPROM/2022/13 from the Generalitat Valenciana.

\appendix
%%%%%%%%%%%%%%%%%%%%%%%%%%%%%%%%%%%%%%%%%%%%%%%%
\section{Coupling terms appearing in the main text}   % Appendix A
\label{sec:appendix_1}
%%%%%%%%%%%%%%%%%%%%%%%%%%%%%%%%%%%%%%%%%%%%%%%%

In this Appendix, we collectively show the coupling terms appearing in the main text. In the order in which they appear, these read
\begin{align}
  {\mathfrak s}_0({\cal P},{\cal A}) 
      =& \frac{1}{2}e^\Phi\big[ 
       - \Phi_{,r} \delta u^{r(1)} \delta u_\phi^{(1)} 
       - \delta u^{r(1)}  \delta u_{\phi,r}^{(1)}
       - \delta u^{\theta(1)} \delta u_{\phi,\theta}^{(1)}\big] 
       + \frac{1}{3} e^\Phi \delta u_\phi^{(1)} \delta u^{\alpha(1)}_{\ ;\alpha},  \\
  {\mathfrak s}_1({\cal P},{\cal A}) 
      =&  \frac{1}{2}e^{-\Phi}\left[ \delta u_\phi^{(1)}\delta u_{r,t}^{(1)}
       + \delta u_r^{(1)} \delta u_{\phi,t}^{(1)} \right],  \\
  {\mathfrak s}_2({\cal P},{\cal A}) 
      =&  \frac{1}{2}e^{-\Phi}\left[\delta u_\phi^{(1)} \delta u_{\theta,t}^{(1)}
       + \delta u_\theta^{(1)}  \delta u_{\phi,t}^{(1)} \right],   \\
  {\mathfrak S}_0({\cal P},{\cal A}) 
      =&  e^{\Phi}\left[ {\mathfrak s}_0({\cal P},{\cal A})  
        - \delta S_{r\phi}^{(1)} \delta u^{r(1)}_{\ ,t} 
        - \delta S_{\theta\phi}^{(1)} \delta u^{\theta(1)}_{\ ,t} 
        - \delta S_{\phi\phi}^{(1)} \delta u^{\phi(1)}_{\ ,t}\right], \\
  {\mathfrak S}_1({\cal P},{\cal A}) 
      =& e^{\Phi}\left[ {\mathfrak s}_1({\cal P},{\cal A}) 
        + \frac{2}{3}\delta S_{r\phi}^{(1)} \delta u^{\alpha(1)}_{\ ;\alpha}
        - \delta u^{r(1)} \delta S_{r\phi,r}^{(1)}
        - \delta u^{\theta(1)} \delta S_{r\phi,\theta}^{(1)} 
        - \delta S_{r\phi}^{(1)} \delta u^{r(1)}_{\ ,r}
        - \delta S_{\theta\phi}^{(1)} \delta u^{\theta(1)}_{\ ,r}
        - \delta S_{\phi\phi}^{(1)} \delta u^{\phi(1)}_{\ ,r}\right],  \\
  {\mathfrak S}_2({\cal P},{\cal A}) 
      =& e^{\Phi}\left[{\mathfrak s}_2({\cal P},{\cal A}) 
        + \frac{2}{3}\delta S_{\theta\phi}^{(1)} \delta u^{\alpha(1)}_{\ ;\alpha}
        - \delta u^{r(1)} \delta S_{\theta\phi,r}^{(1)}
        - \delta u^{\theta(1)} \delta S_{\theta\phi,\theta}^{(1)}
        - \delta S_{r\phi}^{(1)} \delta u^{r(1)}_{\ ,\theta} 
        - \delta S_{\theta\phi}^{(1)} \delta u^{\theta(1)}_{\ ,\theta}
        - \delta S_{\phi\phi}^{(1)} \delta u^{\phi(1)}_{\ ,\theta}\right],    \\
  {\cal S}_F({\cal P},{\cal A})
      =& \frac{(\varepsilon + p)e^{-2\Phi}}{\sin\theta}
        \bigg\{\frac{\delta \varepsilon^{(1)} + \delta p^{(1)}}{\varepsilon + p}\partial_{tt}{\cal Y}^{(1)}
     + r(\partial_t W^{(1)})(\partial_{tr}{\cal Y}^{(1)})
     + (\partial_t V^{(1)})(\partial_{t\theta}{\cal Y}^{(1)})
     + \left(r\partial_{tr} W^{(1)} +\partial_{t\theta} V^{(1)}\right)(\partial_t{\cal Y}^{(1)}) \nonumber \\
     &+\frac{\partial_t(\delta \varepsilon^{(1)} + \delta p^{(1)})}{\varepsilon + p}\partial_t {\cal Y}^{(1)}
     + \left[\frac{\varepsilon' + p'}{\varepsilon + p} - \Phi_{,r}  + \Lambda_{,r} 
     + \frac{5}{r}\right]r(\partial_t W^{(1)})(\partial_t{\cal Y}^{(1)})
     + \frac{2\cos\theta}{\sin\theta}(\partial_t V^{(1)})(\partial_{t}{\cal Y}^{(1)})\bigg\}, \\
  {\cal S}_S({\cal P},{\cal A})
      =& \frac{2\mu}{r^2\sin^2\theta}\bigg[e^{-2\Phi} {\mathfrak S}_0({\cal P},{\cal A}) 
     - e^{-2\Lambda} \left(\partial_r {\cal S}_1({\cal P},{\cal A})
     + \Phi_{,r}' \int_0^t {\cal S}_0({\cal P},{\cal A}) dt 
     + \Phi_{,r} {\cal S}_0({\cal P},{\cal A})\right) \nonumber \\
     &-\frac{1}{r^2} \partial_\theta {\cal S}_2({\cal P},{\cal A})
     - e^{-2\Lambda} \left(\frac{\mu_{,r}}{\mu} +\Phi_{,r} - \Lambda_{,r} + \frac{2}{r}\right)
        \left({\cal S}_1({\cal P},{\cal A})
     + \Phi_{,r} \int_0^t {\cal S}_0({\cal P},{\cal A}) dt\right)
     -\frac{\cos\theta}{r^2\sin\theta}{\cal S}_2({\cal P},{\cal A}) \bigg],
\end{align}
where we note that $\delta u^{r(1)}$ and $\delta u^{\theta(1)}$, which correspond to $W^{(1)}$ and $V^{(1)}$, are of polar parity, while $\delta u_\phi^{(1)}$, which corresponds to ${\cal Y}^{(1)}$, is axial.

%%%%%%%%%%%%%%%%%%%%%%%%%%%%%%%%%%%%%%%%%%%%%%%%
\section{Numerical convergence tests}
\label{sec:appendix_2}
%%%%%%%%%%%%%%%%%%%%%%%%%%%%%%%%%%%%%%%%%%%%%%%%

In this Appendix, we examine the dependence of numerical results on the employed resolution. For this purpose, we study the time evolution of Eq.~(\ref{eq:2nd}), assuming that ${\cal S}_F({\cal P},{\cal A})={\cal S}_S({\cal P},{\cal A})=0$, which is equivalent to the linear perturbation analysis for the torsional oscillations. So, we should confirm that the specific oscillation frequencies obtained from fast Fourier transform (FFTs) of the time-dependent simulation data are close to the frequencies determined via the eigenvalue problem in the no-coupling limit. First, we consider a simulation with initial data given by (say) 
\begin{equation}
  \tilde{\cal Y}^{(2)}(r,\theta) 
     = \frac{1}{3}\left(\frac{r}{R_e}\right)^2\left[1-\left(\frac{r}{R_e}\right)+\left(\frac{r}{R_e}\right)^2  \right]
        \cos\theta,   \label{eq:initial_l2}
\end{equation}
which corresponds to an $\ell=2$-like profile, because $\tilde{\cal Y}$ in the linear analysis is expressed as $\tilde{\cal Y}\sim \partial_\theta P_\ell/\sin\theta$ and $\partial_\theta P_{\ell=2}=3\sin\theta\cos\theta$, even though the radial profile is arbitrary with the radius of the interface between the crust and envelope, $R_e$.

\begin{figure}[tbp]
\begin{center}
\includegraphics[scale=0.5]{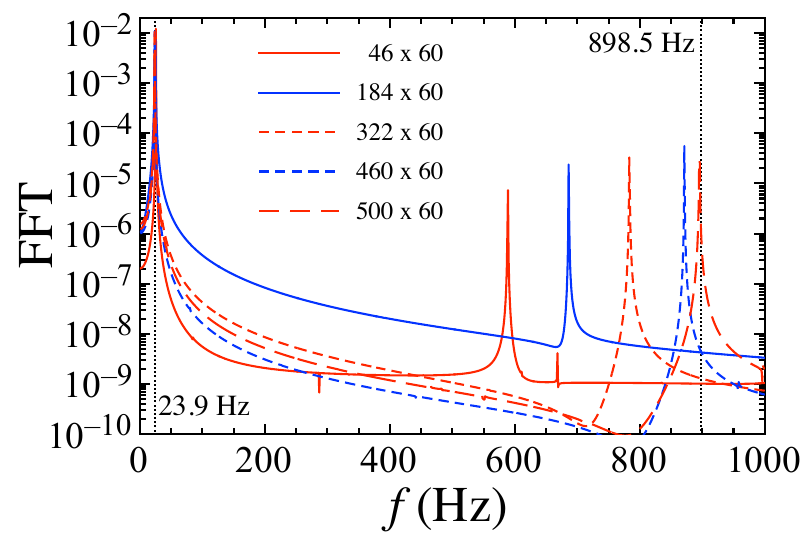}
\end{center}
\caption{%%
The FFT obtained from simulations with various resolutions in the radial direction, but fixed angular resolution $N_\theta=60$. The legend values indicate $N_r\times N_\theta$. For reference, the vertical dotted lines denote the $\ell=2$ frequencies of the fundamental (23.9 Hz) and 1st-overtone (898.5 Hz) torsional oscillations determined via the eigenvalue problem.
}%%
\label{fig:FFT_Nr}
\end{figure}

\begin{figure}[tbp]
\begin{center}
\includegraphics[scale=0.5]{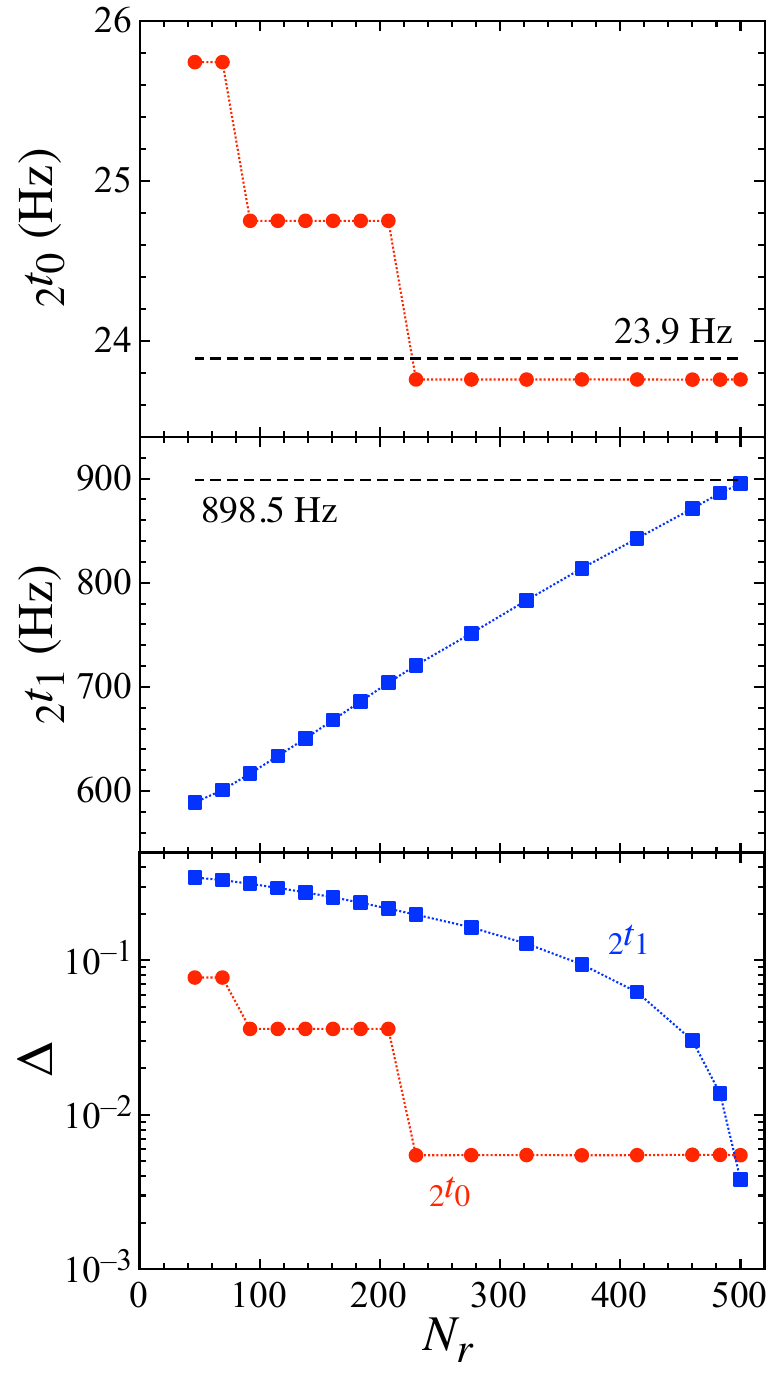}
\end{center}
\caption{%%
Dependence of the frequency of fundamental (top panel) and 1st overtone torsional oscillations (middle panel) obtained from the time evolution on the grid number $N_r$ in the radial direction, where the grid number in the $\theta$ direction is fixed to $N_\theta=60$. The horizontal dashed lines denote the frequencies determined via the eigenvalue problem. In the bottom panel, the relative deviation from the frequencies determined via the eigenvalue problem is shown. 
}%%
\label{fig:dep_Nr}
\end{figure}

Using the initial condition given by Eq.~\eqref{eq:initial_l2}, one can expect that only $\ell=2$ torsional oscillations would be excited because the linear perturbations are not coupled with each other (recalling we have set the sources to zero). In Fig.~\ref{fig:FFT_Nr}, we show the FFT, using the simulation data following evolution for 1 second, by adjusting the radial grid number $N_r$ while keeping the angular one fixed ($N_\theta = 60$). 
In the same figure, we also plot, for reference, the fundamental and 1st overtone frequencies determined via the eigenvalue problem, computed as 23.9 and 898.5 Hz respectively, with dotted vertical lines. We see that the frequency of the 1st overtone determined with the simulation is severely impacted by a lower radial resolution until $N_{r} \gtrsim 500$. To more closely compare the accuracy of the frequencies obtained from the simulation lasting one second, $f_{\rm FFT}$ (symbols), with the eigenvalues, $f_{\rm eigen}$ (dashed lines), the two values are shown as a function of $N_r$ in Fig.~\ref{fig:dep_Nr}; the top and middle panels respectively correspond to the frequencies of the fundamental and 1st overtone torsional oscillations. Since the simulation spans 1 second, our ability to resolve individual frequencies is limited to a band of width $\sim 1$ Hz -- for our purposes, such a Nyquist frequency is already sufficiently small, though we increase the run duration to 10 seconds in the next section. Nevertheless, one observes accuracy improvements with $N_r$ until $N_{r} \sim 500$. To clarify this point, we also check the relative deviation, $\Delta$, of $f_{\rm FFT}$ from $f_{\rm eigen}$, calculated through
\begin{equation}
   \Delta = \frac{|f_{\rm FFT} - f_{\rm eigen}|}{f_{\rm eigen}}. \label{eq:Delta}
\end{equation}
The results are shown in the bottom panel of Fig.~\ref{fig:dep_Nr}. In light of this test, we hereafter adopt $N_r=500$ in this study. 

In a similar way, if one selects an $\ell=3$ mode-like initial condition, i.e., $\tilde{\cal Y}^{(2)}\sim 5\cos^2\theta -1$, and checks the FFT using the simulation data, one can see that only ${}_3t_0$ is excited among the fundamental oscillations, while ${}_2t_0$ is not excited. Or, if one selects the sum of $\ell=2$ and $\ell=3$ mode-like profiles as an initial condition, one can see that ${}_2t_0$ and ${}_3t_0$ are simultaneously excited among the fundamental oscillations.

%%%%%%%%%%%%%%%%%%%%%%%%%%%%%%%%%%%
% Figure 7
%%%%%%%%%%%%%%%%%%%%%%%%%%%%%%%%%%%
%%%%%%%%%%%%%%%%%%%%%%%%%%%%%%%%%%%

%%%%%%%%%%%%%%%%%%%%%%%%%%%%%%%%%%%
% Figure 8
%%%%%%%%%%%%%%%%%%%%%%%%%%%%%%%%%%%
%%%%%%%%%%%%%%%%%%%%%%%%%%%%%%%%%%%

\begin{figure}[tbp]
\begin{center}
\includegraphics[scale=0.5]{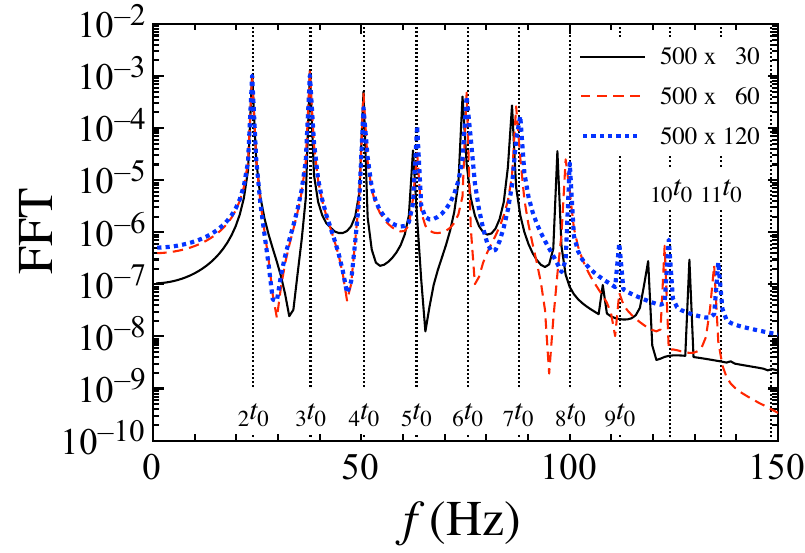}
\end{center}
\caption{%%
Similar to Fig.~\ref{fig:FFT_Nr}, but showing FFTs for simulations with different angular resolution but fixed $N_r=500$. The legends denote $N_r\times N_\theta$. The vertical dotted lines denote the frequencies of fundamental torsional oscillations determined via the eigenvalue problem. Here and elsewhere, ${}_\ell t_n$ denotes the frequency of the $\ell$-th torsional oscillation having $n$ radial nodes in the eigenfunction.
}%%
\label{fig:FFT_Ntheta}
\end{figure}

\begin{figure}[tbp]
\begin{center}
\includegraphics[scale=0.5]{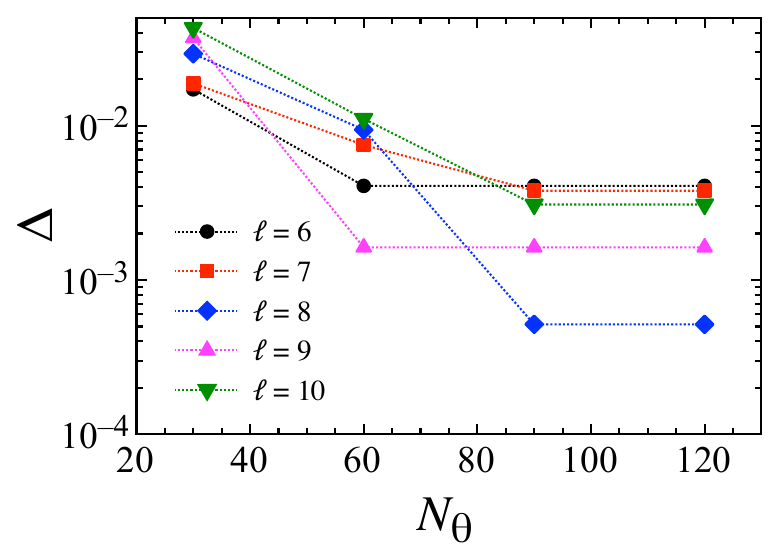}
\end{center}
\caption{%%
Similar to Fig.~\ref{fig:dep_Nr} but showing convergence behavior, as a function of $N_\theta$, of the fundamental torsional oscillation frequencies for $6 \leq \ell \leq 10$, with fixed $N_r=500$. The relative deviation, $\Delta$, of the frequencies obtained by FFT from those determined via the eigenvalue problem is estimated with Eq.~\eqref{eq:Delta}.
}%%
\label{fig:dep_Ntheta}
\end{figure}

Next, we investigate the dependence on $N_\theta$. The oscillations with larger $\ell$ have more nodes in the angular direction. So, to see the dependence on $N_\theta$, it is necessary to study oscillations with large $\ell$. For this purpose, we adopt the initial condition given by
%%%
\begin{equation}
  \tilde{\cal Y}^{(2)}(r,\theta) 
     = \left(\frac{r}{R_e}\right)^2 \left[1  -\left(\frac{r}{R_e}\right) +\left(\frac{r}{R_e}\right)^2\right]\cos^9\theta
        (1 + \cos\theta),   \label{eq:initial_l10}
\end{equation}
which, importantly, does not correspond to the angular profile of any specific mode. Even so, since $\cos^{9}\theta$ and $\cos^{10}\theta$ resemble that of $\ell=10$ and 11 modes, we can expect excitations of modes corresponding to $\ell \leq 11$. Using the initial condition given by Eq.~\eqref{eq:initial_l10} and evolving for 1 second, we compute the FFT shown in Fig.~\ref{fig:FFT_Ntheta} for various angular resolutions: $N_\theta=30$ (solid), 60 (dashed), and 120 (dotted). One can observe that the deviation from the eigenvalue-determined frequencies grows with $\ell$ and is also larger for lower resolution. In Fig.~\ref{fig:dep_Ntheta}, the relative deviation calculated with Eq.~(\ref{eq:Delta}) is shown as a function of $N_\theta$ for $\ell=6-10$. In any case, we find that one reasonably resolve modes reaching $\ell\simeq 10$ with $N_\theta=90$ or 120.

%%%%%%%%%%%%%%%%%%%%%%%%%%%%%%%%%%%
% Figure 9
%%%%%%%%%%%%%%%%%%%%%%%%%%%%%%%%%%%
%%%%%%%%%%%%%%%%%%%%%%%%%%%%%%%%%%%

%%%%%%%%%%%%%%%%%%%%%%%%%%%%%%%%%%%
% Figure 10
%%%%%%%%%%%%%%%%%%%%%%%%%%%%%%%%%%%
%%%%%%%%%%%%%%%%%%%%%%%%%%%%%%%%%%%

%\bibliographystyle{h-physrev} % for PrD
%\bibliography{mybib}
%%%%%%%%%%%%%%%%%%%%%%%%%%%%%%%%%%%%%%%%%%%%%%%%

\end{document}